\documentstyle[prl,aps,epsf]{revtex}
\begin{document}
\newcommand{\bo}{\={o} }
\newcommand{\be}{\begin{equation} }
\newcommand{\ee}{\end{equation} }

\title{Running title: Understanding hierarchical protein evolution}

\title{Understanding hierarchical protein evolution from first
principles}

\author{Nikolay V. Dokholyan\footnote{Email: dokh@wild.harvard.edu}
and Eugene I. Shakhnovich\footnote{Email: eugene@belok.harvard.edu}}

\address{Department of Chemistry, Harvard University, 12
Oxford Street, Cambridge, MA 02138, USA}

%\date{\today ~~~\bf DRAFT}

\maketitle

\begin{abstract}
We propose a model that explains the hierarchical organization of
proteins in fold families. The model, which is based on the
evolutionary selection of proteins by their native state stability,
reproduces patterns of amino acids conserved across protein
families. Due to its dynamic nature, the model sheds light on the
evolutionary time scales. By studying the relaxation of the
correlation function between consecutive mutations at a given position
in proteins, we observe separation of the evolutionary time scales: at
the short time intervals families of proteins with similar sequences
and structures are formed, while at long time intervals the families
of structurally similar proteins that have low sequence similarity are
formed. We discuss the evolutionary implications of our model. We
provide a ``profile'' solution to our model and find agreement between
predicted patterns of conserved amino acids and those actually
observed in nature.
\end{abstract}

\noindent
{\bf Key words: } protein evolution, $Z$-score model, profile solution

\section{Introduction}

Understanding protein evolution still remains a major challenge in
molecular biology
\cite{Levitt76,Chothia92,Finkelstein93,Orengo94,Davidson94,Finkelstein95,Murzin95,Li96,Rost97,Chothia97,Murzin98,Holm98,Buchler00}.
While the mechanisms of mutations in DNA sequences that code for
proteins are known \cite{Alberts94}, the selective fixation of these
mutations in proteins is far from clear. Mutations occurring in DNA
are directly governed by physical-chemical processes and their
fixation is subject to cellular repair mechanisms being able to
preserve nucleotide(s) from modifications. Selection by evolution is
much softer in DNA than in proteins due to a stronger inter-dependence
of the protein structure, function and kinetic properties than that of
DNA. Since mutations may drastically alter the physical, chemical, and
biological properties of proteins, evolution exerts pressure to
preserve those amino acids that play an important role in the folding
kinetics, functionality and stability of proteins.  Our goal is to
understand evolution from the statistical mechanics perspective.

There are several principal facts observed in proteins: {\it (i)} a
protein sequence folds into a unique three-dimensional structure
(there might be exceptions, e.g. prions); {\it (ii)} protein sequences
are {\it selected}, i.e. a randomly chosen polypeptide most likely
aggregates in solution without forming any definite three-dimensional
structure; {\it (iii)} proteins taken from various species and having
sequence identity, $ID$, at least $ID=25$ -- 30\% have similar
three-dimensional structures ({\it native state})
\cite{Holm98,Sander91,Flahetry91,Holmes93,Orengo97,Dodge98,Sanchez00,Perl00}
and are said to belong to the same fold family; {\it (iv)} some pairs
of proteins sharing the same fold have sequence similarity as low as
expected for random sequences $ID\sim 8$ -- 9\%
\cite{Rost97,Holm93,Holm97}; {\it (v)} within the same fold family,
protein sequences have only 3 -- 4\% ``anchored'' amino acids
\cite{Rost97}. A set of proteins that have at least 25\% sequence
similarity and are structurally similar are called homologs. A set of
structurally similar proteins that may have less than 25\% sequence
similarity is called a group of structurally homologous proteins or
analogs. Analogs include several families of homologs and generally
constitute a larger set of proteins than homologs. Known homologs and
analogs are collected in the HSSP \cite{Dodge98} and FSSP
\cite{Holm93} databases respectively and are the subject of our study.

Here we propose a model of evolution ({\it $Z$-score model}) that,
based on facts {\it (i)} and {\it (ii)}, attempts to reproduce the
rest of remaining principal observations {\it (iii) -- (v)} described
above. The $Z$-score model is based on the design of a set of
structurally identical sequences by the $Z$-score minimization
\cite{Shakhnovich93,Abkevich96,Shakhnovich97}. The idea is to find the
similarities in the sequences of such a set and to recover those
residues that are conserved across this set. The protein folding
theory \cite{Bryngelson87,Abkevich98} suggests that $Z$-score
minimization is equivalent to maximizing the energy gap between
misfolded or unfolded conformations and the native state of a
protein. It has been pointed out that such maximization results in
stable and fast-folding proteins \cite{Shakhnovich93,Shakhnovich98}.
Thus, by designing sequences that have the same fold, we attempt to
mimic evolution in diversifying protein sequences for the same fold
family.  In addition, the $Z$-score model is a dynamical model,
i.e. there is an implicit time scale that allows one to follow the
evolution of sequences during the design procedure.  The model is
discussed in detail in Sec.~\ref{s:z} and a profile approximation
to this model is outlined in Sec.~\ref{s:mf}. In Sec.~\ref{s:r} we
show that our view of evolution proposed below is consistent with the
implications of the proposed model. Next we discuss our scenario of
protein evolution.

\begin{figure}[hbt]
 \centerline{ \vbox{ \hbox{\epsfxsize=14.0cm
 \epsfbox{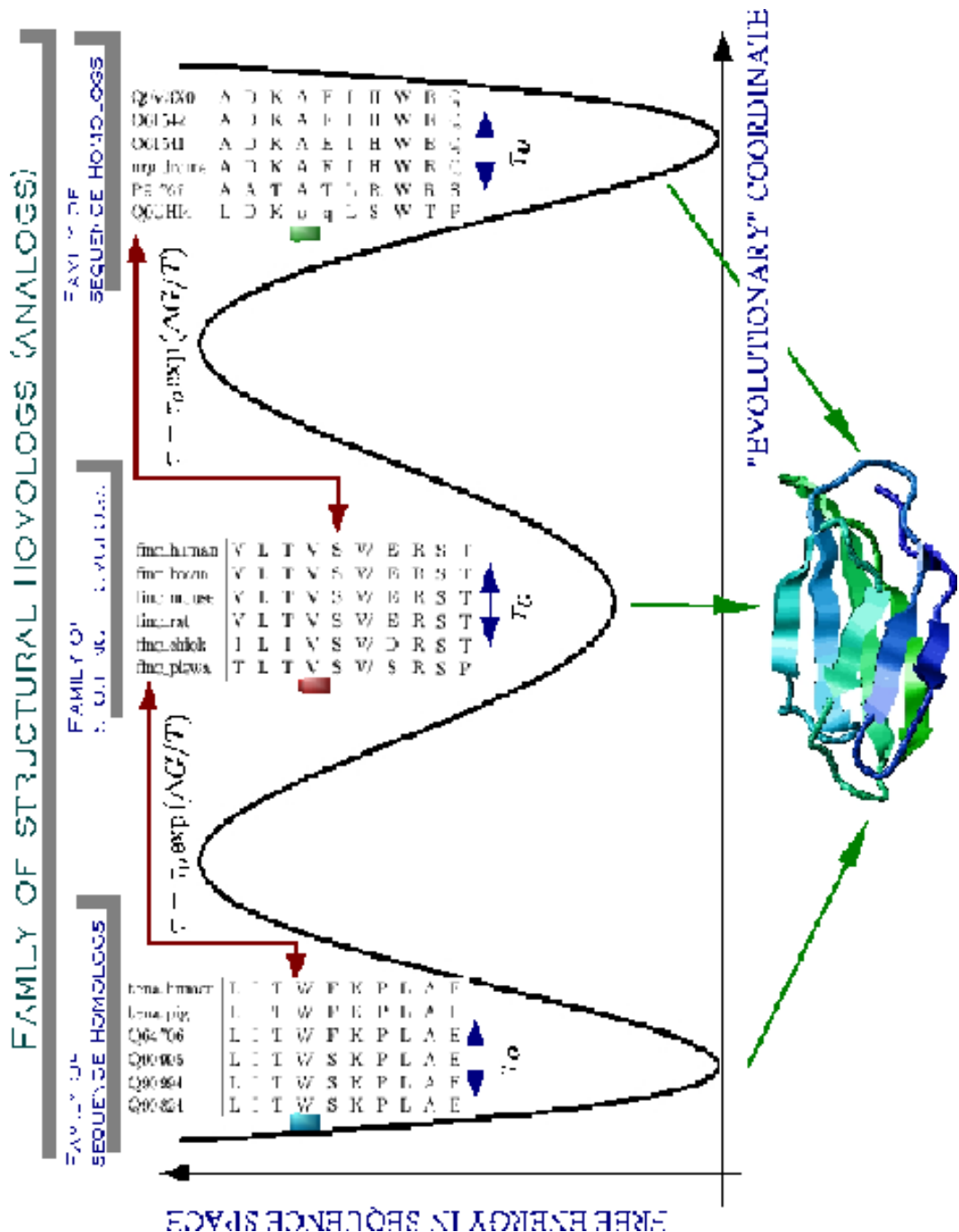}  }}}
 \vspace*{0.5cm}
\caption{A schematic representation of the evolutionary processes that
result in conservation patterns of amino acids. For a given family of
folds, e.g. Ig in this diagram, there are several alternative minima
(3) in the hypothetical free energy landscape in the sequence space as
a function of the ``evolutionary'' reaction coordinate (e.g. time).
Each of these minima are formed by mutations in protein sequences at
time scales, $\tau_o$, that do not alter the protein's
thermodynamically and/or kinetically important sites, forming families
of homologous proteins. Transitions from one minimum to another occur
at time scales $\tau=\tau_0 \exp (\Delta G/T)$. At time scale $\tau$
mutations occur that would alter several amino acids at the important
sites of the proteins in such a way that the protein properties are
not compromised. At time scale $\tau$ the family of analogs is
formed. In three minima we present three families of homologs (1TEN,
1FNF, and 1CFB) each comprised of six homologous proteins. We show 10
positions in the aligned proteins: from 18 to 28. It can be observed
that at position 4 (marked by blocks) in each of the families
presented in the diagram amino acids are conserved within each family
of homologs, but vary between these families. This position
corresponds to position 21 in Ig fold alignment (to 1TEN) and is
conserved (see Figs.~\protect\ref{fig:2}(a)).}
\label{fig:evol}
\end{figure}

We conjecture that hierarchical organization of structurally similar
proteins may be the result of the separation of the evolutionary time
scales, shown schematically on Fig.~\ref{fig:evol}. On a time scale
$\tau_o$, a set of mutations occur that do not affect those amino
acids that play thermodynamical, kinetical and/or functional roles. As
a result, there is little variation in sequences at the important
sites of proteins. If a mutation occurs at the thermodynamically,
kinetically and/or functionally important sites, it usually
substitutes amino acids with close physical properties so that core,
nucleus and/or functional site are not disrupted and the protein folds
into its family fold, is stable in this fold, and its function is
preserved. At this time scale, a family of homologs is born.

Rarely, at time scale $\tau$, correlated mutations occur
\cite{Mirny99,Altschuh88,Thomas96} that modify {\it several} amino
acids at the core, nucleus and/or functional site, so that the
stability and kinetics of proteins are not altered. Such a set of
mutations can drastically modify the sequence of the protein. However,
within the time scale $\tau_o$, a family of homologs is born within
which there is conservation of (already new) amino acids in the
specific (important) sites of homologous proteins. Although there are
alternations in the specific sites of the proteins at the time scale
$\tau$, these sites are more preserved than the rest of the
sequence. The proposed view of protein evolution is consistent with
the observations of hierarchical organization of structurally similar
proteins in families of homologs. Sets of families of homologs are
organized, in turn, in super-families of analogs.

\section{$Z$-score model}
\label{s:z}

\begin{figure}[hbt]
 \centerline{ \vbox{ \hbox{\epsfxsize=10.0cm
 \epsfbox{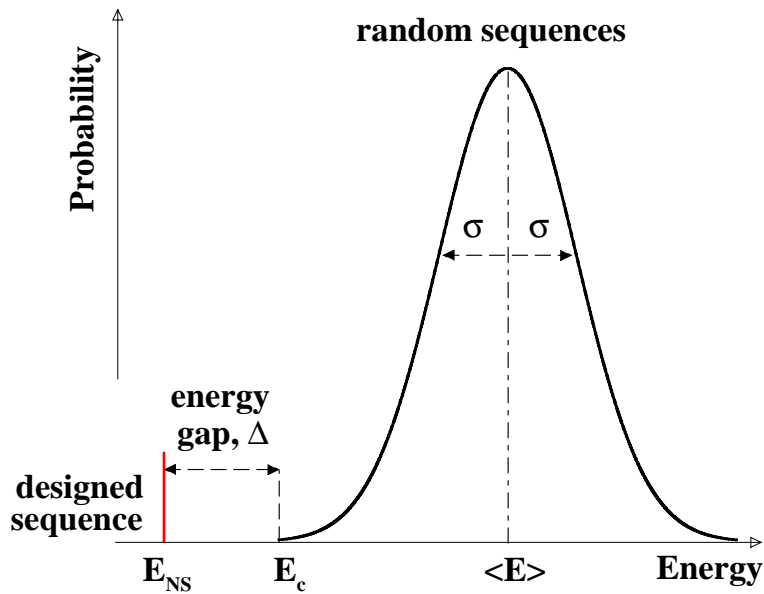}  }}}
 \vspace*{0.5cm}
\caption{A schematic representation of the probability distribution of
energies of the random sequences versus the designed sequence.}
\label{fig:1}
\end{figure}

We start with a random protein amino acid sequence and perform a Monte
Carlo search for the mutation that energetically favors interactions
in such a sequence. The Monte Carlo design algorithm is based on the
minimization of the so-called $Z$-score, defined as
\be
Z = \frac{E_{NS}-\langle E \rangle}{\sigma{(E)}}\, ,
\label{eq:Z}
\ee
which corresponds to the minimization of the energy gap between the
native state, $E_{NS}$, of the selected sequence and the average
energy, $\langle E \rangle$, of structurally unrelated conformations
(decoys) \cite{Gutin98a,Buchler99,Mirny00}. $\sigma(E)$ is the
standard deviation of energies of all decoys (see Fig.~\ref{fig:1}).

Since $Z$-score minimization is equivalent to maximizing the energy
gap between misfolded or unfolded conformations and the native state
of the protein \cite{Bryngelson87,Abkevich98,Gutin98a}, such
maximization results in stable and fast-folding proteins. The energy
gap must be ``significant'', meaning that $E_{NS}$ must deviate from
$\langle E \rangle$ by many standard deviations $\sigma$: $E_{NS} \ll
\langle E \rangle - \sigma$. Many researchers have pointed out (see,
e.~g., review \cite{Shakhnovich98}) that minimization of the $Z$-score
corresponds to the stabilization of the protein in its native state.

The design proceeds as follows: {\it (i)} we select an amino acid
$\sigma_i$ at a random position $1\leq i\leq N$; {\it (ii)} we
substitute this amino acid by $\sigma_i'$ with probability p,
\be
p = \left\{ \begin{array}{ll}
         1, & \mbox{if}\, \delta Z<0\\
         \exp(-\delta Z/T_{des}), &   \mbox{if}\, \delta Z >0\, ,
            \end{array} \label{eq:p}
\right.
\ee
where $\delta Z = Z(\sigma_i') - Z(\sigma)$ is the difference between
the $Z$-scores of the mutated and the original proteins. We design
each of $N_s=100$ sequences by running the simulations for $N_m$ Monte
Carlo steps at some design temperature, $T_{des}$.

Computation of $\langle E\rangle$ and $\sigma(E)$ is straightforward:
\be
\langle E\rangle = \frac{1}{2} \sum_{i\neq j} U(\sigma_i, \sigma_j)
f_{ij}\, ,
\label{eq:Eav}
\ee
and 
\be
\sigma^2(E) = \langle (E - \langle E\rangle)^2\rangle =
\frac{1}{2} \sum_{i\neq j} f_{ij} (1 - f_{ij}) U^2(\sigma_i, \sigma_j)
+ {\cal O}(f^2_{ij})\, .
\label{eq:sigE}
\ee
where $f_{ij}$ is the frequency of a contact between monomers $i$ and
$j$ in a set of decoys, i.~e.
\be
f_{ij} = \langle \Delta_{ij}\rangle\, .
\label{eq:fij}
\ee

We estimate frequencies of contacts by making two assumptions about
the set of decoys: {\it (1)} the distribution, $P(\ell=|i-j|; i, j)$
of the contact distances, $\ell=|i-j|$, between various amino acids at
the positions $i$ and $j$ is universal among globular proteins; and
{\it (2)} the actual frequency of contacts between various amino
acids, $i$ and $j$, is only a function of the absolute value of the
length of contacts, $|i-j|$, and is equal to the distribution of the
contact lengths, i. e.
\be
f_{ij} = f_{|i-j|} = f(\ell)\, . 
\ee
The distribution $P(\ell)$ is then
\be
P(\ell) = {f(\ell)\over \sum_{\ell=1}^{N} f(\ell)}\, .\label{eq:P-l}
\ee
Both assumptions, {\it (1)} and {\it (2)}, are motivated by the fact
that the variety of protein structures known to date samples
adequately the conformational space of proteins under study and the
variety is large enough that all the information about the secondary
structure peculiarities of individual proteins are averaged out.

In order to estimate frequencies, $f_{ij}$, according to
Eq.~(\ref{eq:P-l}) we compute the distribution of contacts of length
$\ell=|i-j|$ in the ensemble of approximately $10^3$ representative
globular proteins in Protein Data Bank (PDB)
\cite{Bernstein77,Abola87}. The distribution, shown in
Fig.~\ref{fig:freq}, is obtained using $C_{\beta}$-representation of
proteins. The contacts are defined by Eq.~(\ref{eq:Delta}).

The estimation of contact frequencies $f_{ij}$ is one of the key
ingredients to protein design. An alternative approach based on
sampling of homopolymer conformations appears to be less efficient, so
we omit it in the present study. Nevertheless, due to its importance
and possible potential for other studies, we discuss this approach in
Appendix \ref{ap:freq}.

After we obtain $N_s$ number of designed sequences, we compute the
probability of an amino acid $\sigma_k$ to be in $k$th position,
$P_Z(\sigma_k)$, as the frequency of occurrence of this amino acid,
\be
P_Z(\sigma_k) = N(\sigma_k)/N_s\, ,
\ee
where $N(\sigma_k)$ is the total number of occurrences of an amino
acid $\sigma_k$ at the position $k$. Next, using Eq.~(\ref{eq:s}) we
compute the sequence entropy, $S_Z(k)$.

\section{Profile solution}
\label{s:mf}

We develop a profile solution to the $Z$-score model that provides
a rationale for conservatism patterns caused by selection for
stability. Our solution is of {\em equilibrium} evolution that
maintains stability and other properties achieved at an earlier,
prebiotic stage. To this end we propose that stability selection
accepts only those mutations that keep energy of the native protein,
$E$, below a certain threshold $E_c$ necessary to maintain an energy
gap \cite{Finkelstein95,Shakhnovich93,Ramanathan94,Sali94}. The
requirement to maintain an energy threshold for the viable sequences
makes the equilibrium ensemble of sequences analogous to a
microcanonical ensemble. In analogy with statistical mechanics, a more
convenient and realistic description of the sequence ensemble is a
canonical ensemble, whereby strict requirements on energy of the
native state is replaced by a ``soft'' evolutionary pressure that
allows energy fluctuations from sequence to sequence but makes
sequences with high energy in the native state unlikely.  In the
canonical ensemble of sequences, the probability of finding a
particular sequence, $\{\sigma\}$, in the ensemble follows the
Boltzmann distribution
\cite{Finkelstein95,Shakhnovich93,Ramanathan94,Pande95}
\be
P(\{\sigma\})=\frac{\exp(-E\{\sigma\}/T)}{Z}\, ,
\ee
where $T$ is the effective temperature of the canonical ensemble of
sequences that serves as a measure of evolutionary pressure and $Z =
\sum_{\{\sigma\}} \exp{( -E\{\sigma\}/T)} $ is the partition function
taken in sequence space.

Next, we apply a profile approximation that replaces all
multiparticle interactions between amino acids with interaction of
each amino acid with an effective field $\Phi$ acting on this amino
acid from the rest of the protein, so that each amino acid experiences
the exact field of its neighbors. This approximation presents
$P(\{\sigma\})$ in a multiplicative form as $\prod_{k=1}^{N}
p(\sigma_k)$ of probabilities to find an amino acid $\sigma$ at
position $k$ \cite{Saven97}. $p(\sigma_k)$ also obeys Boltzmann
statistics
\be
p(\sigma_k) = \frac{\exp(-\Phi(\sigma_k)/T)}{\sum_{\sigma}
\exp(-\Phi(\sigma_k)/T)}\, .
\label{eq:pmf}
\ee
The profile potential $\Phi(\sigma_k)$ is the effective
potential energy between amino acid $\sigma_k$ and all amino acids
interacting with it, i.~e.
\be
\Phi(\sigma_k)=\sum_{i\neq k }^{N} U(\sigma_k, \sigma_i)\Delta_{ki}\, .
\label{eq:phi}
\ee
The potential $\Phi$ is similar in spirit to the protein profile
introduced by Bowie et al. \cite{Bowie91} to identify protein
sequences that fold into a specific 3D structure.

For each member, $m$, of the fold family (FSSP database \cite{Holm93})
presented in Fig.~\ref{fig:evol}, we compute the profile
probability, $p_m(\sigma_k)$, using Eq.~(\ref{eq:pmf}). This
probability, $p_m(\sigma_k)$, for each fold family member corresponds
to the frequency of amino acids, $\sigma_k$, at positions, $k$, for a
given family of homologs. Then, we compute the average profile
probability over all members of the fold family,
\be
p_{P}(\sigma_k) = \frac{1}{N_s} \sum_{m=1}^{N_s} p_m(\sigma_k)\, .
\label{eq:pmf1}
\ee
This quantity corresponds to the $P_{acr}(\sigma_k)$ presented in
\cite{Mirny99}.  Eqs. (\ref{eq:pmf}) --- (\ref{eq:pmf1}), along with
the properly selected energy function, $U$, make it possible to predict
probabilities of all amino acid types and sequence entropy $S_{P}(k)$
at each position $k$
\be
S_{P}(k)=- \sum_{\sigma}p_{P}(\sigma_k)\ln p_{P}(\sigma_k)\,
\label{eq:SP}
\ee
from the native structure of a protein. The summation is taken over
all possible values of $\sigma$.

If stability selection is a factor in the evolution of proteins and
our model captures it, then we should observe a correlation between
the predicted profile based sequence entropies, $S_{P}(k)$, and
actual sequence entropies $S_{acr}(k)$ in real proteins.  Thus, the
question is: ``Can we find such $T$, so that the predicted
conservatism profile $S_{P}(k)$ matches the real one $S_{acr}(k)$?''

By varying the values of the temperature $T$ in the range $0.1\leq
T\leq 4.0$, we minimize the distance, $D^2\equiv\sum_{k=1}^N
(S_{P}(k) - S_{acr}(k))^2$, between the predicted and observed
conservatism profiles. We exclude from this sum such positions in
structurally aligned sequences that have more than 50\% gaps in the
structural (FSSP) alignment. We denote by $T_{sel}$ the temperature
that minimizes $D$.

The proposed profile solution has a dual role. On one hand, it
allows us to understand the selective temperature scale, $T_{sel}$, which
is the measure of evolutionary optimization. On the other hand, the
correlation coefficient between $S_{P}(k)$ and $S_{acr}(k)$ does not
vary strongly in the range of $T_{sel}$ from 0.19 to 0.34, thus,
allowing one to use the effective temperature of $T_{sel}=0.25$ to
predict the actual conservatism profiles of proteins (see
Table~\ref{t:2}).

\section{Results and Discussion}
\label{s:r}

\begin{figure}[hbt]
 \centerline{ \vbox{ \hbox{\epsfxsize=10.0cm
 \epsfbox{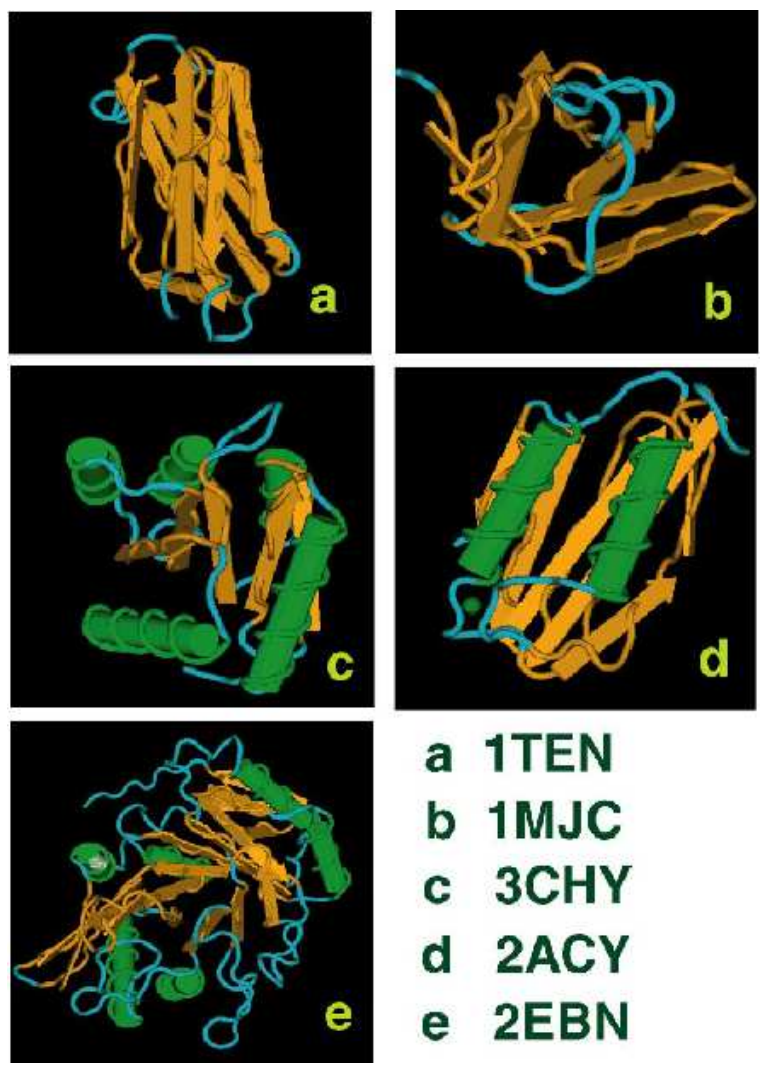}  }}}
 \vspace*{0.5cm}
\caption{Three dimensional structures of the representative proteins
of the five folds under study (Ig, OB, R, $\alpha/\beta$-P and TIM
folds): (a) 1TEN protein, (b) 1MJC, (c) 3CHY, (d) 2ACY, and (e) 2EBN.}
\label{fig:5folds}
\end{figure}

We study five folds: Immunoglobulin fold (Ig), Oligonucleotide-binding
fold (OB), Rossman fold (R), $\alpha/\beta$-plait ($\alpha/\beta$-P),
and TIM-barrel fold (TIM). The three-dimensional structures of the
representative proteins of these five folds are shown in
Fig.~\ref{fig:5folds}: (a) Tenascin (Third Fibronectin Type III
Repeat), pdb:1TEN; (b) Major Cold Shock Protein 7.4 (Cspa (Cs 7.4)) of
{\it Escherichia Coli}, pdb:1MJC; (c) chemotactic protein CheY, pdb:3CHY;
(d) Acyl-Phosphatase (Common Type) From Bovine Testis, pdb:2ACY; (e)
Endo-Beta-N-Acetylglucosaminidase F1, pdb:2EBN. We compute the
correlation coefficient \cite{Press89} between values of $S_{P}(k)$,
obtained at $T_{sel}$, and $S_{acr}(k)$ for all five folds. The
results are summarized in Table~\ref{t:2}. The plots of $S_{P}(k)$
and $S_{acr}(k)$ versus $k$, as well as their scatter plots, are shown
in Figs.~\ref{fig:2}--\ref{fig:6}.

\subsection{$Z$-score model}
\label{ss:z}

We find that correlation between $S_{Z}(k)$ and $S_{acr}(k)$ strongly
depends on the number of mutations, $N_m$, we introduce during design
of a protein. This fact is in accord with our view (see
Fig.~\ref{fig:evol}) of protein evolution. On a short time scale,
$\tau_o\sim 10^2$ Monte Carlo steps, mutations rarely alter amino
acids with specific important properties such as participation in
stabilization of proteins and/or in the nucleation processes in
folding kinetics of proteins. These mutations diversify the family,
$m$, of homologs, ${\cal M}_h^m$.  On a larger scale, $\tau \gg
\tau_o$, correlated mutations \cite{Mirny99,Altschuh88,Thomas96}
modify core and/or nucleus site(s) of the proteins without
compromising their stability, folding rates and function(s). Thus, at
the time scale $\tau$ evolution moves from one family of homologs to
another, diversifying the underlying family of analogs, ${\cal M}_a$,
$\bigcup\limits_m {\cal M}_h^m \subseteq {\cal M}_a$. The ensemble of
analogs is still much smaller than the ensemble, ${\cal M}_o$, of all
possible sequences (${\cal M}_a \subseteq {\cal M}_o$), which is of
the size $6^N$ (in a 6-letter alphabet) --- for $N=100$ residue
protein this number is of the order of $10^{78}$. These results are in
agreement with the theoretical predictions
\cite{Finkelstein93,Bryngelson87,Shakhnovich98,Shakhnovich89,Shakhnovich90,Pande97}
that there is a large number (of the order $e^{1.9 N}$
\cite{Shakhnovich98}) of fast folding sequences with a given native
structure and pronounced stability gaps $\Delta = E_{NS} - \langle E
\rangle$.

It is important that for the small number of mutations we find
correlation between entropies of the designed sequence, $S_{Z}(k)$,
and the empirically observed one, $S_{acr}(k)$. This correlation
depends on the input random number, indicating that the selected
sequences constitute a family of homologs, ${\cal M}_h^m$, that is
closer or more distant to an original sequence family of homologs,
${\cal M}_h^0$ (both ${\cal M}_h^m$ and ${\cal M}_h^0$ belong to a
given family of analogs, ${\cal M}_a$). Here we present the results
for the selected ensembles of the designed sequences, ${\cal M}_h^m$,
after being optimized during $N_m$ mutations. More important than the
correlation between $S_{Z}(k)$ and $S_{acr}(k)$, we find that the
profiles of $S_{Z}(k)$ and $S_{acr}(k)$ are in visible concert with
each other.

\begin{figure}[hbt]
 \centerline{ \vbox{ \hbox{\epsfxsize=10.0cm
 \epsfbox{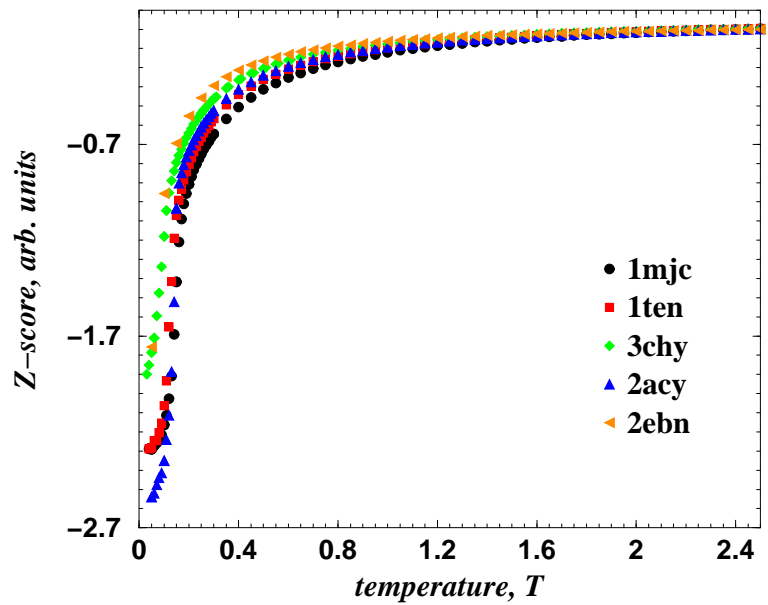}  }}}
 \vspace*{0.5cm}
\caption{The temperature dependence of the $Z$-score obtained after
$10^5$ Monte Carlo steps and averaged over 100 design ``trajectories''
for five representative proteins (1MJC, 1TEN, 3CHY, 2ACY, and 2EBN) of
the five folds studied. Due to normalization of the contacts'
frequencies extracted from PDB database, the scales of values of $Z$
are different from the actual $Z$-score with correctly normalized
$Z-score$.}
\label{fig:z}
\end{figure}

The temperature dependence of the $Z$-score exhibits a sharp
transition at $T = T_c\approx 0.25$ (Fig.~\ref{fig:z}) for all studied
proteins. Above $T_c$, protein design results in unstable sequences,
while at temperatures much lower than $T_c$ many of the residues
``freeze'' in their original states. Thus, we select $T_c$ as our
design temperature.

\subsection{Degree of divergence of sequences}
\label{ss:div}

To assess the degree of similarity or divergence of sequences in the
course of $Z$-score design at various time scales, we compute the
distribution of hamming distances at these time scales. The hamming
distance, $Hd (\{\sigma\}^{(1)}, \{\sigma\}^{(2)})$, between two
sequences, $\{\sigma\}^{(1)}$ and $\{\sigma\}^{(2)}$, is defined as
the number of distinct amino acids at equal positions in these two
sequences divided by the length of the sequences, $N$:
\be
Hd (\{\sigma\}^{(1)}, \{\sigma\}^{(2)}) = {1\over N} \sum_{i=1}^{N}
\Big[1-\delta(\sigma_k^{(1)} - \sigma_k^{(2)})\Big]\, .
\ee
Hamming distance has a simple interpretation --- it is the degree of
divergence between two sequences: when $Hd$ is equal to 1, the
sequences have no amino acids in common, when $Hd$ is equal to 0, the
sequences are exactly the same.

\begin{figure}[hbt]
 \centerline{
 \vbox{ \hbox{\epsfxsize=10.0cm \epsfbox{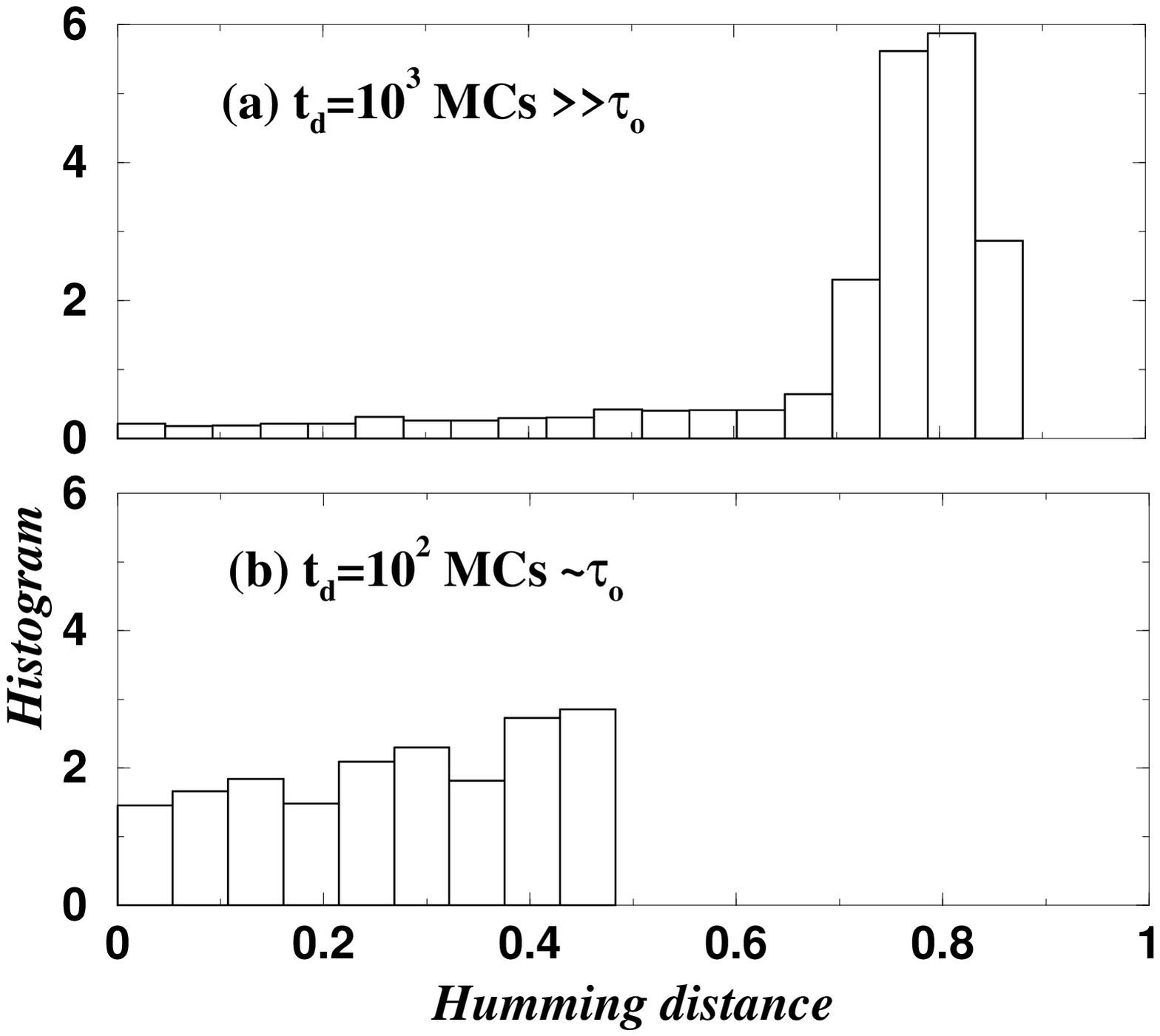}  }}
 \vbox{ \hbox{\epsfxsize=10.0cm \epsfbox{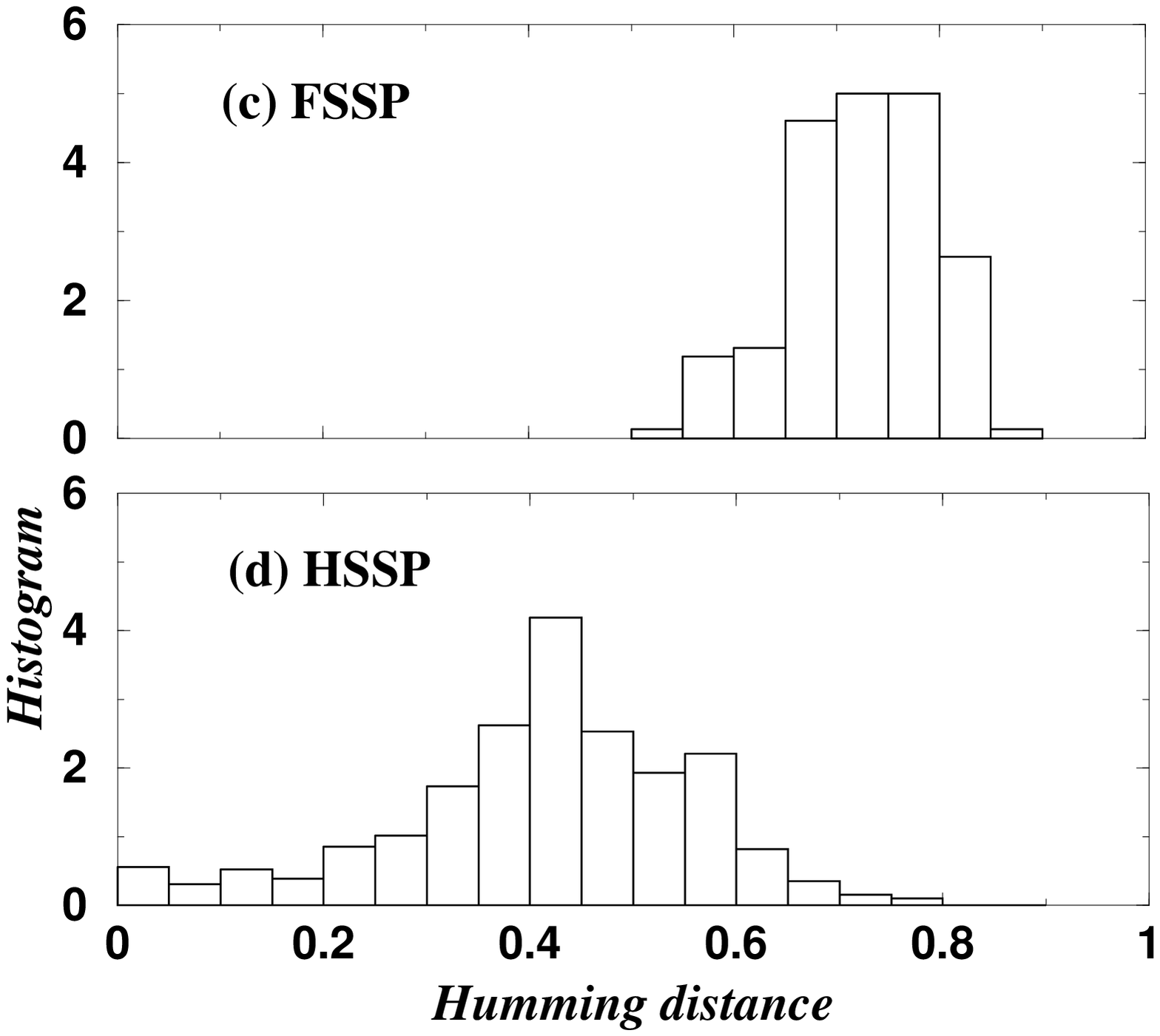}  }}
 }
 \vspace*{0.5cm}
\caption{The histograms of hamming distances for 1MJC family between
all designed sequences for two design times (a) $t_d =10^3 \gg \tau_o$
and (b) $t_d=10^2 \sim \tau_o$ Monte Carlo steps.  The histograms of
hamming distances for 1MJC family of actual protein sequences: (c)
analogs, taken from the FSSP database, and (d) homologs, taken from
the HSSP database. In the computation of the histograms in case (b)
we omit all sequences with sequence similarity less than $ID = 55\%$
to mimic sequence collection in the HSSP database. The threshold
sequence similarity, $ID$, is chosen so that the hamming distance
histogram derived from the actual sequences in HSSP database
(d). All histograms are normalized to unit area.}
\label{fig:8}
\end{figure}

We compute the distribution of hamming distances between all designed
sequences for two design times (a) $t_d =10^3 \gg \tau_o$
(Fig.~\ref{fig:8}(a)) and (b) $t_d=10^2 \sim \tau_o$ Monte Carlo steps
(Fig.~\ref{fig:8}(b)), where $\tau_o$ is a characteristic time
scale. We use 1MJC family of homologs (OB fold) as an example
throughout this subsection. We also performed similar analysis with
other folds and the results are qualitatively the same (not shown).

In the computation of the distribution in case (b) we omit all
sequences with sequence similarity less than $ID = 55\%$ to mimic
sequence collection in the HSSP database. This threshold sequence
similarity, $ID$, is chosen so that the hamming distance distribution
derived from the actual sequences in the HSSP database
(Fig.~\ref{fig:8}(d)) is similar to ours. Given that we use a
six-letter alphabet, the correspondence between $ID$ used in HSSP and
our $ID$ is not well defined. Because in (b) we select only sequences
with minimal threshold similarity, $ID$, there are no events with
$Hd>0.55$ in Fig.~\ref{fig:8}(b). In addition, the events with low $Hd
\rightarrow 0$ are overrepresented in our simulations since we do not
account for additional pressure due to function or kinetics that
exists in real protein sequences. Therefore, the distribution in our
simulations (Fig.~\ref{fig:8}(b)) has a more pronounced tail $Hd
\rightarrow 0$ than that in real proteins (Fig.~\ref{fig:8}(d)).

At the long time scales (a) we find that most of the sequences are
divergent from each other with average $\langle Hd \rangle \approx
0.7$. We observe the same result by computing the distribution of the
hamming distances between all analogs belonging to the OB fold family
present in the FSSP database (Fig.~\ref{fig:8}(c)). The only
difference between simulated (Fig.~\ref{fig:8}(a)) and observed
(Fig.~\ref{fig:8}(c)) distributions of hamming distances is the tail
present in the simulated distribution corresponding to the sequences
with a significant degree of similarity. This tail is due to the fact
that we compare all sequence with all sequences, thus, effectively
including similar sequences in our histogram. In the FSSP database,
on the other hand, only distant sequences are present so that the
tail corresponding to the close sequences in Fig.~\ref{fig:8}(c) is
absent.

The distributions of hamming distances in protein families are in
qualitative agreement with those observed in simulations and with our
picture of hierarchical protein evolution. At short time scales
sequences are not strongly separated from each other forming families
of homologs, while at long time scales, a family of analogs is formed,
comprised of strongly separated sequences but structurally similar
proteins.

\subsection{Determination of the family formation time scale}

To quantify our observation of evolutionary time scales separation, we
compute the relaxation times of the correlation function (at each protein
position, $k$) in the course of $Z$-score design defined as
\be
C_k(\tau) = {1\over t_d N_s} \sum_{\alpha = 1}^{N_s} \int_{0}^{t_d}
\chi_k^{(\alpha)}(t,\tau) dt = \langle\langle
\chi_k(\tau)\rangle\rangle_{t_d, N_s} \, ,
\label{eq:C}
\ee
where $\langle\langle\dots\rangle\rangle_{t_d, N_s}$ denotes average
over simulation design time, $t_d$, and the number, $N_s$, of initial
sequences. $\chi_k^{(\alpha)}(t,\tau)$ is a boolean indicator of
whether an amino acid $\sigma_k(t+\tau)$ at position $k$ at time
$t+\tau$ is the same as the amino acid at time $\sigma_k(t)$ at the
same position at time $t$:
\be
\chi_k^{(\alpha)}(t,\tau) = \left\{ \begin{array}{ll}
                   1, & \sigma_k(t) = \sigma_k(t+\tau) \\
                   0, & \sigma_k(t) \neq \sigma_k(t+\tau) \, .
                  \end{array} \label{eq:chi}
\right.
\ee
$C_k(\tau)$ measures the probability that a mutation does not occur at
the position $k$ in time $\tau$. This function for most equilibrium
systems decays exponentially,
\be
C(\tau) \sim \exp (-\tau/\tau_o)\, ,
\ee
where $\tau_o$ is the relaxation time that is the average mutation
time between subsequent mutations.

\begin{figure}[hbt]
 \centerline{
 \vbox{ \hbox{\epsfxsize=10.0cm \epsfbox{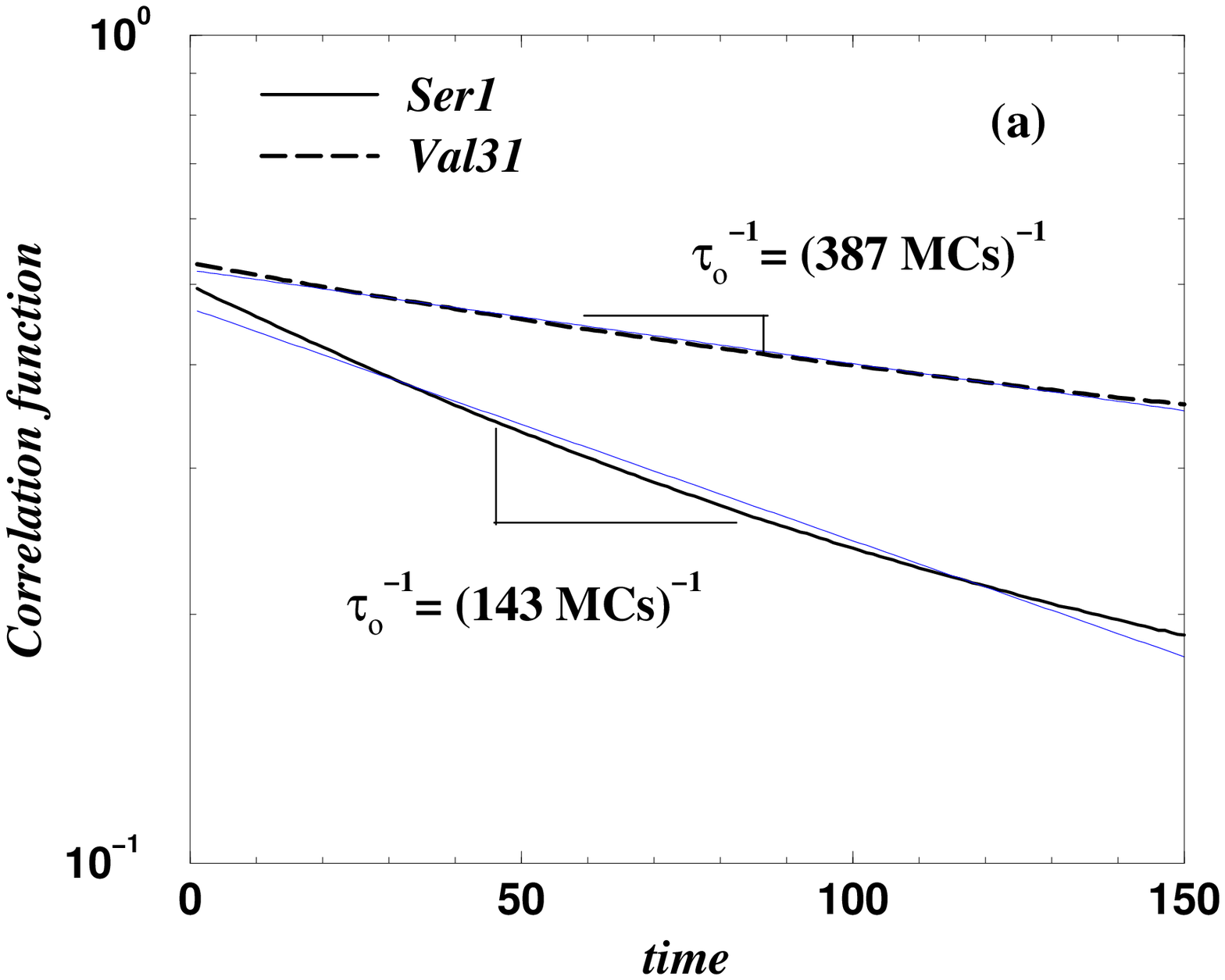}  }}
 \vbox{ \hbox{\epsfxsize=10.0cm \epsfbox{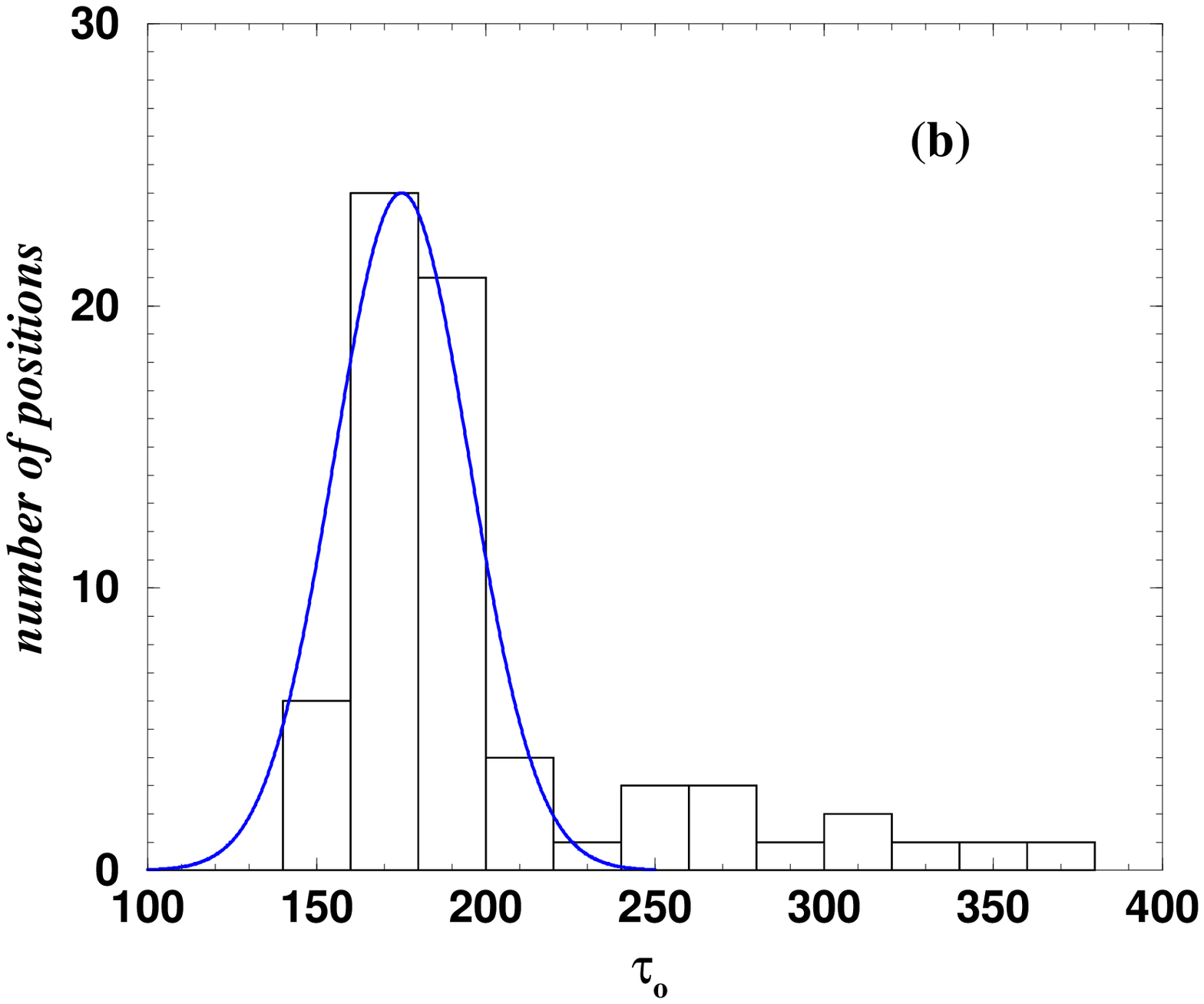}  }}
 }
 \vspace*{0.5cm}
\caption{(a) Plot of the correlation functions versus time for two
position in 1MJC, Ser1 and Val31, obtained in the course of $Z$-score
design of $10^3$ sequences for $10^3$ Monte Carlo steps. In
semilogarithmic scale $C_{k=1}(\tau)$ and $C_{k=31}(\tau)$ are
straight lines with slopes $\tau_o = 143$ and $\tau_o = 387$ Monte
Carlo steps correspondingly. (b) The histogram of the relaxation times
$\tau_o$ for all positions in 1MJC obtained in the course of $Z$-score
design of $10^3$ sequences for $10^3$ Monte Carlo steps. The histogram
is well fit by a Gaussian function in the region $100 < \tau_o < 250$
(solid line). The long tail that strongly deviates from the Gaussian
distribution (over seven standard deviations) indicates the presence
of the conserved positions in the course of design.}
\label{fig:7}
\end{figure}

We also find that the correlation function computed for $N_s=10^3$ and for
$t_d=10^3$ decays exponentially (see Fig~\ref{fig:7}) and relaxation
times $\tau_o$ depend strongly on the positions of the amino acids
under consideration. For example, the relaxation of the correlation
functions for positions 1 (Ser in 1MJC) and 31 (Val) in 1MJC design
vary by almost a factor of two: $\tau_o(\mbox{Ser1})=143$ and
$\tau_o(\mbox{Val31}) = 387$ Monte Carlo steps. The fact that
$\tau_o(\mbox{Ser1})$ is more than two times larger than
$\tau_o(\mbox{Val31})$ indicates that Ser1 is likely to mutate more
than twice in the time-span of a single Val31 mutation.

In addition, the distribution of the relaxation times (see
Fig.~\ref{fig:8}) exhibits a pronounced peak at $\tau_o=170$ Monte
Carlo steps, indicating that for most protein positions relaxation
occurs with this typical relaxation time. The relaxation times found
from the correlation function analysis are in agreement with our
observations in Secs.~\ref{ss:z} and \ref{ss:div}. The long
non-gaussian tail in the histogram of the relaxation times also
suggests the presence of the conserved positions. In fact, this tail,
composed of the conserved positions, strongly deviates from the rest
of the distribution, which is well approximated by a Gaussian
distribution.

\subsection{Rates of amino acid substitutions and conservatism}
\label{ss:freq}

A number of authors suggested
\cite{Ptitsyn98,Mirny98,Ptitsyn99,Dokholyan00xx} that study of the
conserved amino acids in families of structurally similar proteins can
shed light on the functionally, kinetically and thermodynamically
important amino acids in proteins. The basic belief behind the
majority of such studies is that evolution optimizes, to a certain
extent, the properties of proteins so that they become more stable
and have better folding and functional properties. Here we use the
``optimization'' hypothesis of molecular evolution to understand the
universe of protein sequences by implication of molecular evolution.
The link between conserved amino acids and their role in proteins has
been widely studied
\cite{Ptitsyn99,Krebs83,Lesk80,Hollecker83,Plaxco97,Nishimura00}.

A recent study of Mirny and Shakhnovich \cite{Mirny99} identified the
presence of universally conserved amino acids across the families of
proteins sharing the same fold. These conserved residues have been
linked to protein stability, kinetic properties or function. Various
experiments
\cite{Krebs83,Lorch99,Schindler98,Lopez96,Russell98,Welch94,Bellsolell96,Wilcock98,Villegas98,vanNuland98a,vanNuland98b}
have identified some of the conserved residues to have predicted
specific roles.

Direct evidence of the relationship between conservatism and the
physical properties of amino acids can be accessed by calculating the
rates of amino acid substitutions in the course of the $Z$-score
design. By comparing mutational rates at various positions of the
proteins, we attempt to reconstruct the conservatism of these positions
across the family of analogous proteins. Starting with the sequence of a
representative protein of a given fold we perform $Z$-score design for
$t_d = 10^8$ Monte Carlo steps. The substitution rates are defined as
\be
R(k) \equiv {N_m(k) \over \hat t_d} = {N\over t_d} \sum_{t=1}^{t_d} \Big[
1- \delta \big( \sigma_k(t) - \sigma_k(t-1) \big) \Big] \, ,
\label{eq:R}
\ee
where $N_m(k)$ is the number of mutations that occurred at the position
$k$, $\delta(x)$ is a Kronecker function, equal to 1 if $x=0$ and 0
otherwise, $\sigma_k(t)$ is an amino acid $\sigma$ at the position $k$
at time $t$, and $\hat t_d = t_d/N$ is the average number of attempted
mutations per position in a protein. Thus, $R(k)$ from the
Eq.~(\ref{eq:R}) is inversely proportional to the average time between
subsequent substitutions of amino acids at the position $k$; the lower
the $R(k)$ the longer the amino acid at the position $k$ remains
unchanged and, therefore, the more conservative is this position in
the course of design.

\begin{figure}[hbt]
 \centerline{ \vbox{ \hbox{\epsfxsize=10.0cm
 \epsfbox{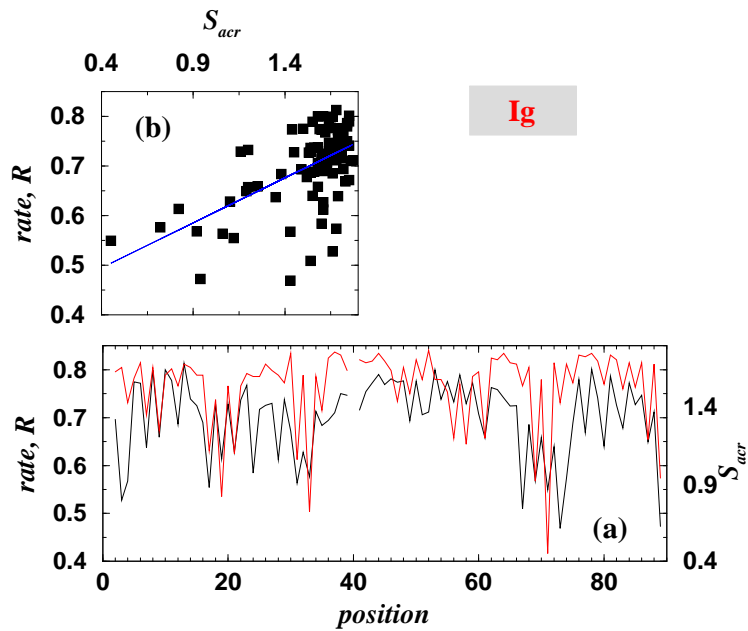}  }}}
 \vspace*{0.5cm}
\caption{(a) The values $R(k)$ (black line) and $S_{acr}(k)$ (red
line) for all positions, $k$, for the Ig-fold. The lower the values of
$R(k)$ the more conservative amino acids are at these positions.  (b)
The scatter plot of $R(k)$ versus observed $S_{acr}(k)$. The linear
regression correlation coefficients are shown in
Table~\protect\ref{t:2}. The blue line is the linear regression
approximation. In both parts (a) and (b) rates are multiplied by the
length of the representative protein.}
\label{fig:2b}
\end{figure}

\begin{figure}[hbt]
 \centerline{ \vbox{ \hbox{\epsfxsize=10.0cm
 \epsfbox{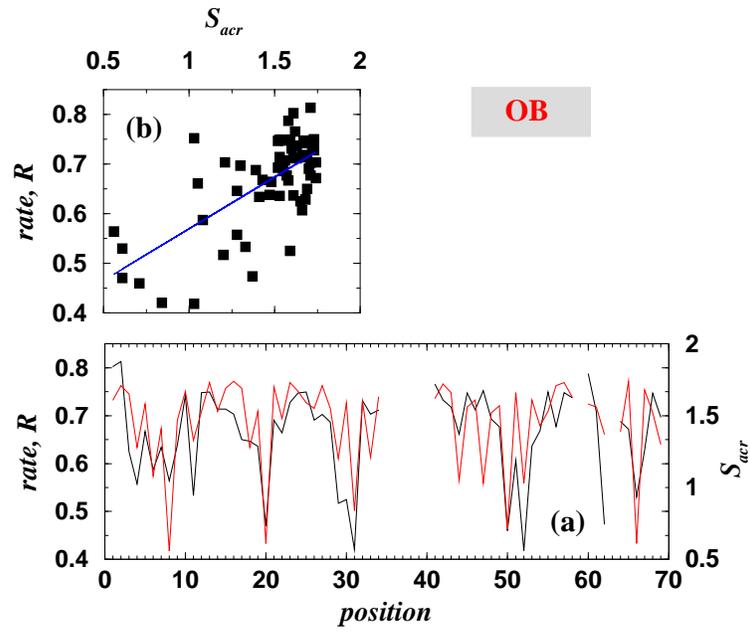}  }}}
 \vspace*{0.5cm}
\caption{(a) --- (c) The same as Fig.\protect\ref{fig:2b} but for the
OB-fold.}
\label{fig:3b}
\end{figure}

\begin{figure}[hbt]
 \centerline{ \vbox{ \hbox{\epsfxsize=10.0cm
 \epsfbox{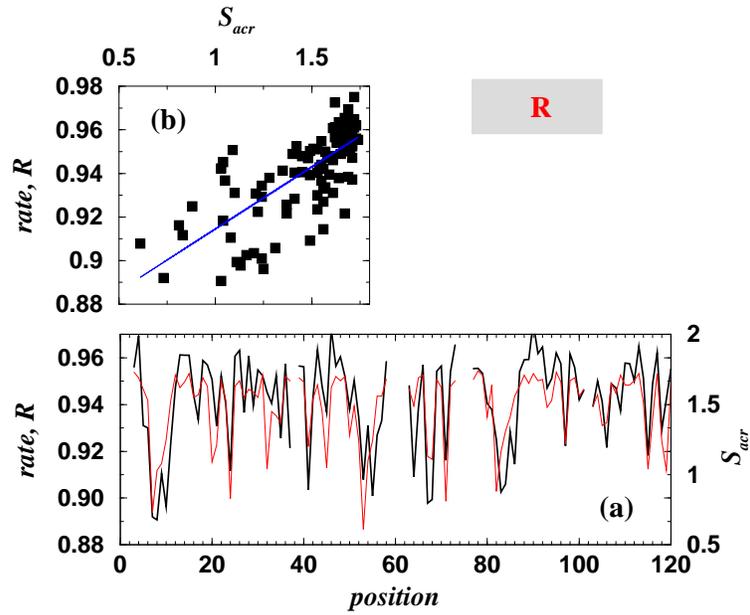}  }}}
 \vspace*{0.5cm}
\caption{(a) --- (c) The same as Fig.\protect\ref{fig:2b} but for
R-fold.}
\label{fig:4b}
\end{figure}

\begin{figure}[hbt]
 \centerline{ \vbox{ \hbox{\epsfxsize=10.0cm
 \epsfbox{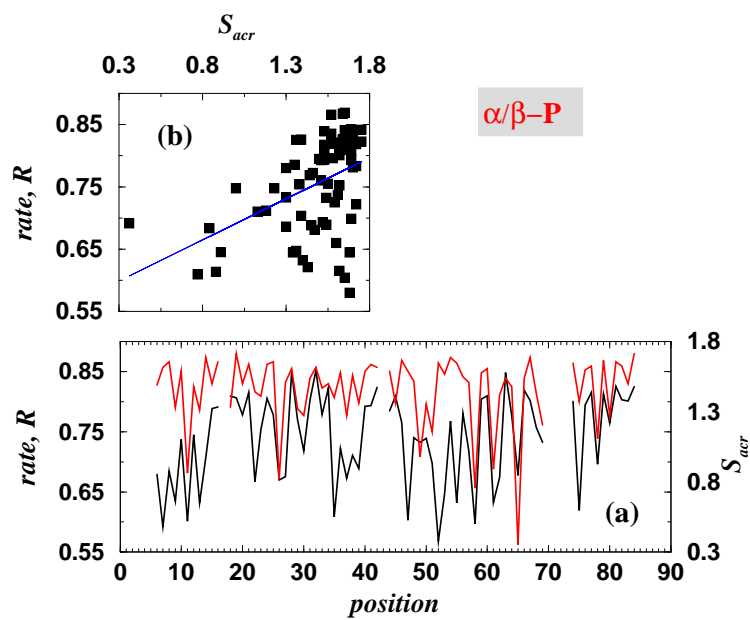}  }}}
 \vspace*{0.5cm}
\caption{(a) --- (c) The same as Fig.\protect\ref{fig:2b} but for the
$\alpha/\beta$-P-fold.}
\label{fig:5b}
\end{figure}

\begin{figure}[hbt]
 \centerline{ \vbox{ \hbox{\epsfxsize=10.0cm
 \epsfbox{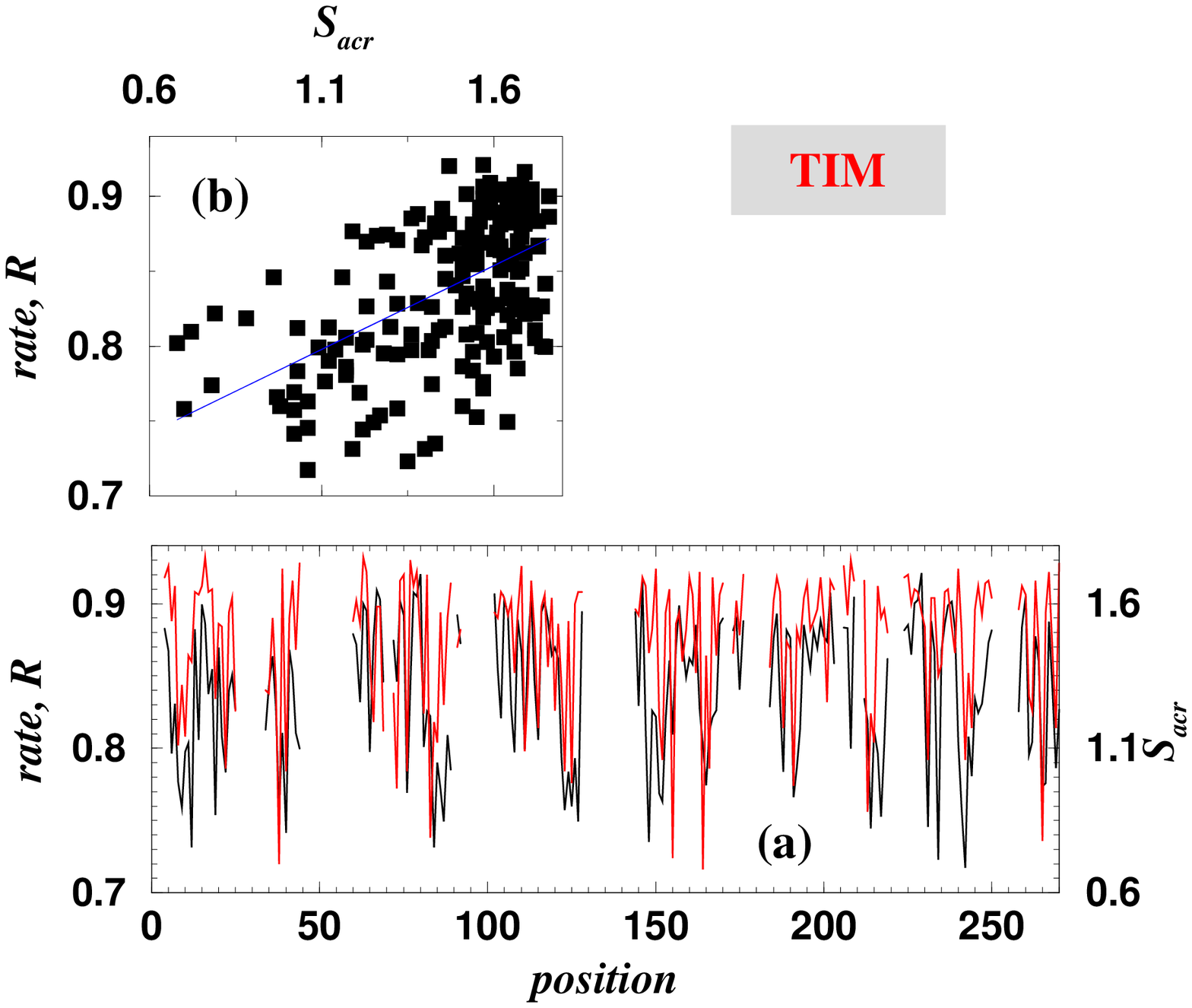}  }}}
 \vspace*{0.5cm}
\caption{(a) --- (c) The same as Fig.\protect\ref{fig:2b} but for
TIM-fold.}
\label{fig:6b}
\end{figure}

We find that the rates of substitutions, $R(k)$, correlate with the
conservatism patterns, $S_{acr}$ (see Figs.~\ref{fig:2b} --
\ref{fig:6b}). Since there is no obvious relation between $R(k)$ and
$S_{acr}$, and, moreover, there is no reason to assume linear relation
between these quantities, the linear regression has only an
illustrative meaning of the correlations observed between $R(k)$ and
$S_{acr}$ (see Figs.~\ref{fig:2b} -- \ref{fig:6b}(b) and
Table~\ref{t:2}). Despite the likely lack of linear relation between
the rates and the entropy, the correlation observed based on the
assumption of linear dependence between $R(k)$ and $S_{acr}$ is
feasible.

The computation of mutational rates, $R(k)$, does not involve the
tuning of any parameters. We can choose any non-zero temperature,
given that the total number of Monte Carlo steps, $t_d$, in the course
of design is large enough to obtain statistically significant values
of $R(k)$. We also find that at $T_{des}=0.25$ the data for $R(k)$ is
identical after $10^7$ Monte Carlo steps to that after $10^8$ Monte
Carlo steps, so the values of $R(k)$ are statistically significant.

Interestingly, the fastest rates are at most two times faster than the
slowest rates. Such variability of rates might be due to the
variability in physical properties of amino acids. It has been shown
\cite{Li97a} that there are only two principal eigenvalues of
Miyazawa-Jernigan energy matrix \cite{Miyazawa85,Miyazawa96} and the
remaining eigenvalues are close to each other. Such a ``degeneracy'' in
eigenvalues accounts for the similarities in physical properties of
amino acids.

Another possible reason for such a wide range of ratevariability is
the absence of the side chains in our model. The side chains are an
additional factor that slow down the rates because of the frustrations
caused by the multiple side chain conformations \cite{Kussell01} and,
possibly, increase the range of rate variability. Despite all the
artifacts of our model, the correlation between $R(k)$ and $S_{acr}$
is significant, which indicates that the model does qualitatively
capture the evolutionary selection of proteins.

\subsection{Profile solution}

The correlation between $S_{P}(k)$ and $S_{acr}(k)$ is remarkable for
all five folds and indicates that our profile approximation is able
to select the conserved amino acids in protein fold families and
properly describe the formation of families on the short time scales
(Table~\ref{t:2}). It is fully expected that the correlation
coefficient is smaller than 1. The reason for this is that computation
of $S_{P}(k)$ takes into account evolutionary selection for stability
only and it does not take into account possible additional pressure to
optimize kinetic or functional properties.

The additional evolutionary pressure due to the kinetic or functional
importance of amino acids results in pronounced deviations of $S_{P}$
from $S_{acr}$ for a few amino acids that may be kinetically or
functionally important. A number of amino acids whose conservatism is
much greater than predicted by our model form a group of ``outliers''
from otherwise very close correspondence between $S_{P}$ and
$S_{acr}$. To demonstrate that some of those amino acids are important
for folding kinetics and, as such, they can be under additional
evolutionary pressure, we color data points on the $S_{P}$ versus
$S_{acr}$ scatter plot according to the range of $\phi$-values
\cite{Itzhaki95} that the corresponding amino acids fall into.  The
thermodynamic and kinetic roles of individual amino acids were studied
extensively {\it (i)} by Hamill et al.  \cite{Hamill00} for the TNfn3
(1TEN) protein, {\it (ii)} by L\'opez-Hern\'andez and Serrano
\cite{Lopez96} for the chemotactic protein (CheY, pdb:3CHY), and {\it
(iii)} by Chiti et al. \cite{Chiti99} for muscle acylphosphatase (AcP,
pdb:2ACY).

\begin{figure}[hbt]
 \centerline{ \vbox{ \hbox{\epsfxsize=10.0cm
 \epsfbox{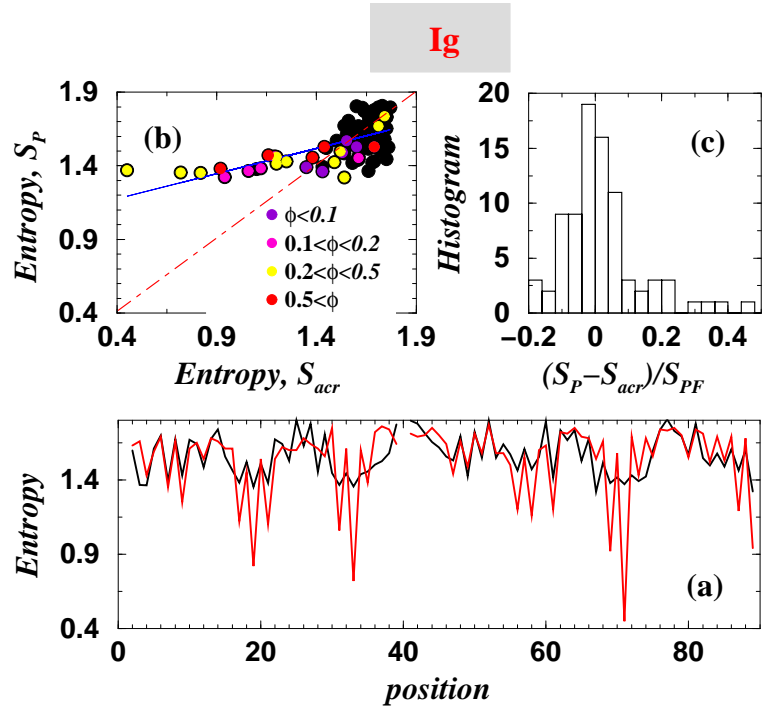}  }}}
 \vspace*{0.5cm}
\caption{(a) The values $S_{P}(k)$ (black line) and $S_{acr}(k)$ (red
line) for all positions, $k$, for the Ig-fold. The lower the values of
$S_{P}(k)$ the more conservative amino acids are at these positions.
(b) The scatter plot of predicted $S_{P}(k)$ versus observed
$S_{acr}(k)$. The linear regression correlation coefficients are shown
in Table~\protect\ref{t:2}. The blue line is the linear regression
that has a slope different than 1 (red line), corresponding to the
$S_{P}(k) = S_{acr}(k)$ relation. (c) The histogram of the relative
differences between $S_{P}(k)$ and $S_{acr}(k)$.  In (b) we assign
colors to data points corresponding to amino acids with the specific
range of $\phi$-values \protect\cite{Hamill00}: red, if $0.5<\phi<1$,
yellow, if $0.2<\phi<0.5$, magenta, if $0.1<\phi<0.2$, violet if
$\phi<0.1$, and black if $\phi$-values are not determined.}
\label{fig:2}
\end{figure}

\begin{figure}[hbt]
 \centerline{ \vbox{ \hbox{\epsfxsize=10.0cm
 \epsfbox{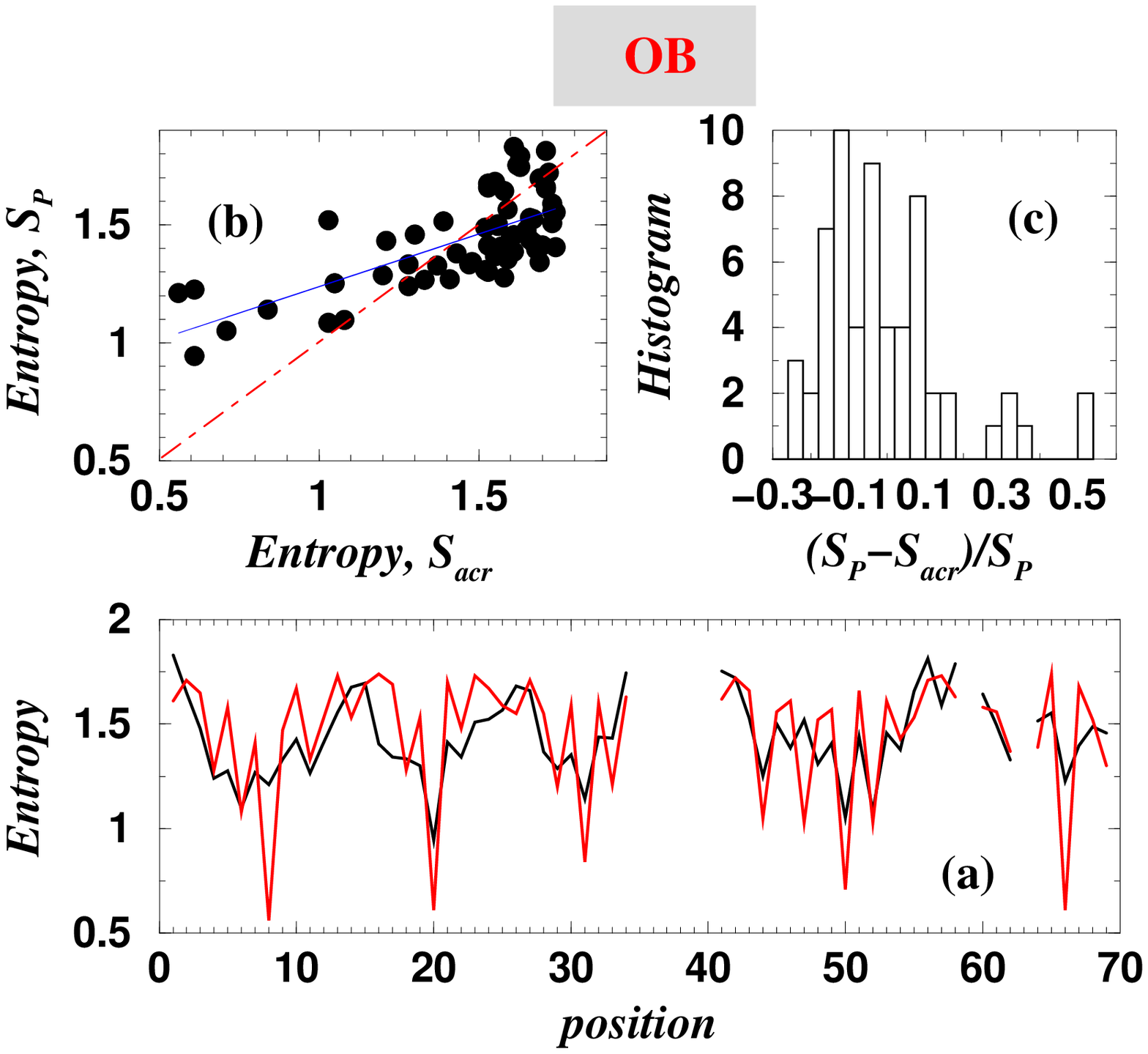}  }}}
 \vspace*{0.5cm}
\caption{(a) --- (c) The same as Fig.\protect\ref{fig:2} but for the
OB-fold.}
\label{fig:3}
\end{figure}

\begin{figure}[hbt]
 \centerline{ \vbox{ \hbox{\epsfxsize=10.0cm
 \epsfbox{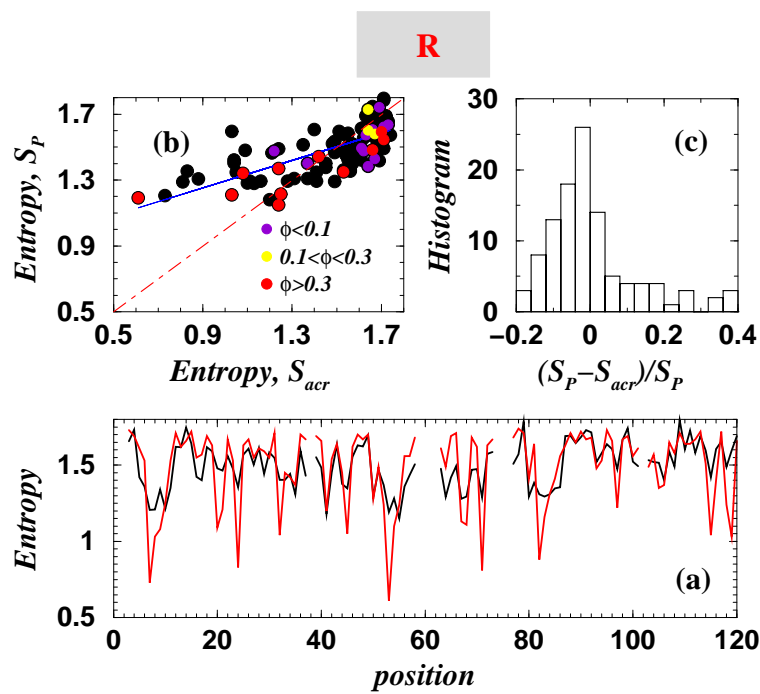}  }}}
 \vspace*{0.5cm}
\caption{(a) --- (c) The same as Fig.\protect\ref{fig:1} but for
R-fold.  In (b) we assign colors to data points corresponding to amino
acids with the specific range of $\phi$-values \protect\cite{Lopez96}:
red, if $0.3<\phi<1$, yellow, if $0.1<\phi<0.3$, violet if $\phi<0.1$,
and black if $\phi$-values are not determined.}
\label{fig:4}
\end{figure}

\begin{figure}[hbt]
 \centerline{ \vbox{ \hbox{\epsfxsize=10.0cm
 \epsfbox{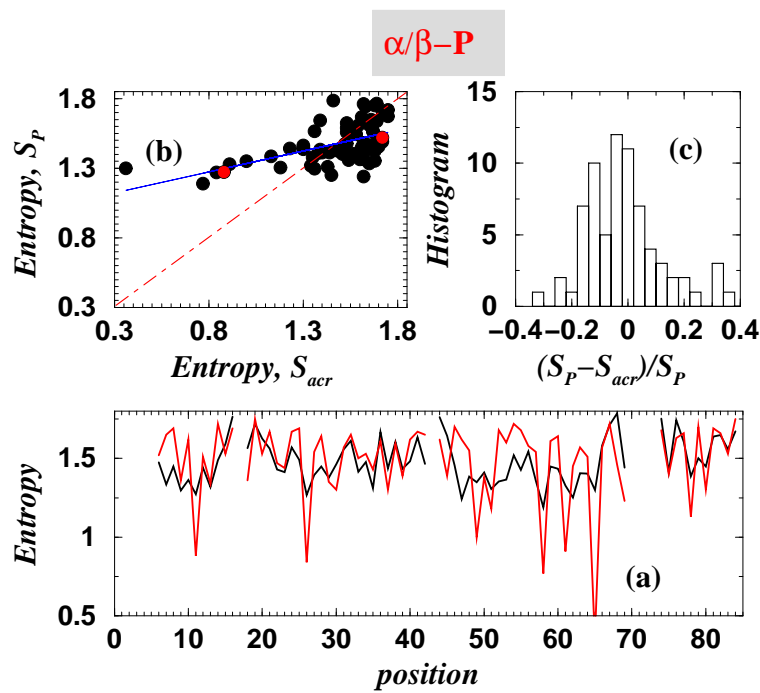}  }}}
 \vspace*{0.5cm}
\caption{(a) --- (c) The same as Fig.\protect\ref{fig:2} but for the
$\alpha/\beta-P$-fold. In (b) we color red (2 out 3) nucleic amino acids,
Tyr11 and Pro54, \protect\cite{Chiti99}.}
\label{fig:5}
\end{figure}

\begin{figure}[hbt]
 \centerline{ \vbox{ \hbox{\epsfxsize=10.0cm
 \epsfbox{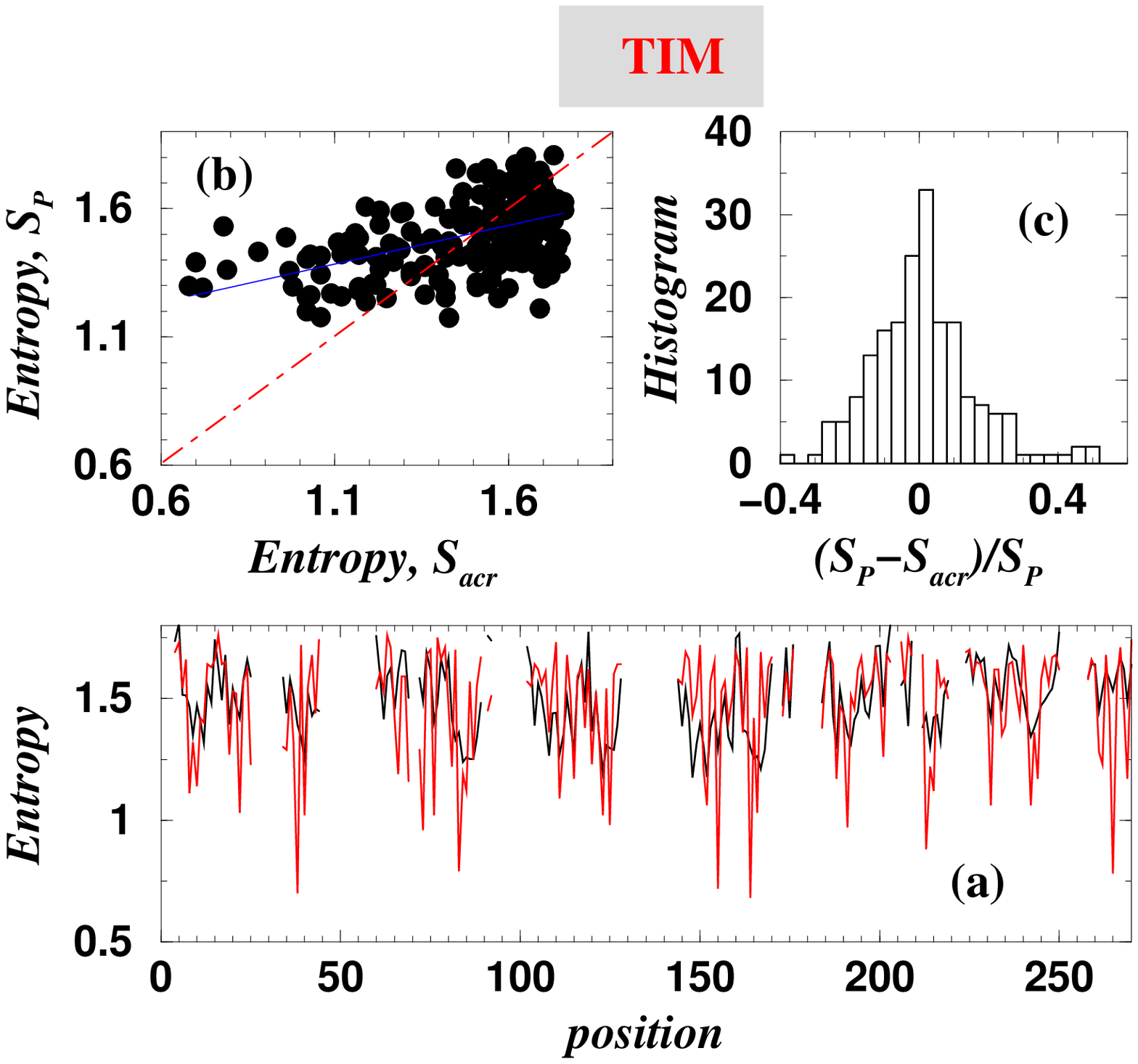}  }}}
 \vspace*{0.5cm}
\caption{(a) --- (c) The same as Fig.\protect\ref{fig:2} but for the
TIM-fold.}
\label{fig:6}
\end{figure}

We use the $\phi$-values for individual amino acids obtained in
\cite{Lopez96,Hamill00}. We observe that {\it (i)} for TNfn3 protein
most of the points on Fig.~\ref{fig:2}(b) that belong to the outlier
group have $\phi$-values ranging from 0.2 to 1; {\it (ii)} for CheY
protein most of the points (for which $\phi$-values are known) on
Fig.~\ref{fig:4}(b) that belong to the outlier have $\phi$-values
ranging from 0.3 to 1; and {\it (iii)} for AcP protein, one nucleic
amino acid, Tyr11, is strongly conserved, more than predicted by the
profile solution, while the second amino acid, Pro54, belonging to the
nucleus \cite{Chiti99} does not appear to be conserved. The third
nucleus amino acid, Phe94, in AcP protein is excluded from our
analysis due to the lack of data at position 94. The discrepancy of
the Pro54 conservatism and its kinetic role may be attributed to the
poor statistical significance of $S_{acr}(k)$ calculation at this
position. Figs.~\ref{fig:2}(b), \ref{fig:4}(b), and \ref{fig:5}(b)
demonstrate that the presence of additional evolutionary pressure due
to the kinetic importance of amino acids results in stronger
conservatism of specific positions than predicted by profile solution.

It has been conjectured (see e.g. \cite{Rost97}) that on average only
3 -- 4\% of residues are ``anchor residues'', i.e. those that are more
significantly conserved than the rest of the residues. In fact, this
observation is supported by the $S_{acr}(k)$ profile of the sequences
and their profile estimates $S_{P}(k)$ (see Figs.~\ref{fig:2}(a) --
\ref{fig:6}(a)). These 3 -- 4\% of ``anchor residues'' are the
principal ``gates'' to the structure/kinetics of a given family of
proteins. For example, it has been shown \cite{Dokholyan00a} that the
number of residues that belong to the nucleus of a model protein is
about 5\%; we expect the same low percentage of residues that
determine the kinetics of real proteins. The number of key residues
that form a functional site is also a small fraction of the total
number of residues in proteins.

In order to demonstrate the statistical significance of the outliers'
kinetic importance, we show that the number of sites with high values
of $\phi$ found among the outliers is larger than that expected if
such sites were randomly distributed across all values of $S_{P}$.
For {\it Tenascin}, the total number of residues is $N_{tot}=89$, the
number of sites with $\phi>0.2$ is $N_{tot}(\phi>0.2)=17$, the number
of outliers is $N_{out}=13$, and the expected number of sites with
$\phi>0.2$ among outliers is $N^{exp}_{out}(\phi>0.2) =
N_{tot}(\phi>0.2) {N_{out} / N_{tot}} \approx 2.5$. The observed number
of sites with $\phi>0.2$ among outliers is $N_{out}(\phi>0.2)=8$,
which is over three times more than expected. A similar estimate for
$\phi>0.5$ gives $N^{exp}_{out}(\phi>0.5)=0.75$ and $N_{out}(\phi>0.5)
= 2$, which is nearly three times more than expected.  For {\it CheY},
the total number of residues is $N_{tot}=128$, the number of sites
with $\phi>0.3$ is $N_{tot}(\phi>0.3)=11$, the number of outliers:
$N_{out} = 22$, and the expected number of sites with $\phi>0.3$ among
outliers is $N^{exp}_{out}(\phi>0.3)=N_{tot}(\phi>0.3) {N_{out}/
N_{tot}} \approx 1.9$. The observed number of sites with $\phi>0.3$ among
outliers is $N_{out}(\phi>0.3)=4$, which is over two times more than
expected. These crude estimates demonstrate that outliers have, in
fact, a higher than expected number of residues with pronounced kinetic
role, hinting towards an additional evolutionary pressure exerted on
kinetically important amino acids.

\subsection{Convergent or divergent evolution?}

It has been a long-standing question
\cite{Rost97,Murzin98,Zhang97,Artymiuk97,Bryant97} whether the
presently known proteins have evolved from a smaller family of
prebiotic proteins (``divergent'' evolution scenario) or whether they
evolved from ancestors with distant homology and due to thermodynamic,
kinetic, and functional pressure exerted by evolution they {\it
converged} to structurally similar proteins (``convergent'' evolution
scenario). The model of evolution, proposed in this work does not rule
out any of these scenario. However, the similarity in distribution of
hamming distances in the family of homologous proteins produced by our
model to that taken from nature is striking (Fig.~\ref{fig:8}),
serving as a hint in favor of divergent evolution.

An important argument favoring diverging evolution is that a function
of a protein is strongly susceptible to protein structure
\cite{Murzin98,Alberts94}. So, if a protein were to change the
structure in the course of evolution, it would affect its
functionality (there are, of course, possible exceptions).
Functionless genes have little chance of surviving in cells, so these
proteins would most probably be eliminated. However, Murzin
\cite{Murzin98} proposed a way for functionless protein to survive by
fusion with another functional protein and evolving already as a unit
to a multifunctional protein. One of the most prominent examples are
the DNA polymerases that are composed of similar domains with
different sequence composition \cite{Wang97,Doublie97,Kiefer97}.

If we set aside multi-domain proteins, the fact that there is a
limited amount of folds ($ < 1,000$ according to Chothia
\cite{Chothia92} or $<7,920$ according to Orengo \cite{Orengo94}) has
been extensively used to favor convergent evolution
\cite{Finkelstein93,Finkelstein95,Li96,Chothia97,Buchler00}. Li et
al. \cite{Li96} used the designability principle to show from full
enumeration of lattice protein models that the number of members of a
fold family depends on the stability gap $\Delta$. This dependence
means that many unrelated sequences search in the course of evolution
for the stable conformations and as soon as they reach a basin of a
certain fold with large enough energy gap they stay within that
basin. The scenario proposed by
\cite{Finkelstein93,Finkelstein95,Li96} also explains why various fold
families are unequally populated \cite{Orengo94} --- the number of
family members depends on the energy gap. The more pronounced the
energy gap is the more mutations such a fold can tolerate. Buchler and
Goldstein \cite{Buchler00} argued that the energy gap depends on the
number of non-local contacts of a given fold.

There are several questions about that scenario. First, it is not
clear if nature exploits all possible folds \cite{Orengo94}: even
though there are only 1,000 folds \cite{Chothia92}, it is possible
that nature simply does not need more of them. Second, Chothia and
Gerstein \cite{Chothia97} argued that the restriction on the
divergence of proteins from one another does not come from a stability
requirement (which is of course important) but from the separation of
the mutated residue from the active site. Thus, the extent of sequence
divergence is inversely proportional to that of protein
function(s). The experiments by Gassner et al. \cite{Gassner96} and
Axe et al. \cite{Axe96} support the arguments of Chothia and Gerstein
\cite{Chothia97}. In these experiments substitution of the several
amino acids in the hydrophobic core of the T4 lysozyme conserved to a
certain extent the function and the structure of the protein. Third,
we note that there is a limited amount of types of chemical elements
that are part of the ligand structures that are bound by the active
site. Thus, we expect that there are groups of evolutionary unrelated
proteins with similar binding sites and the structures. In fact, there
are examples of proteins sharing the same site, also called a
``super-site''. For example, both transforming protein p21H-RAS-1
fragment (pdb:1CTQA) and chemotactic protein 3CHY have similar binding
sites, the root-mean-square deviation of one protein from another is
3.2\r A while there are only 13 identical residues (i.e. $ID < 10
\%$).  Interestingly, the active site of 1CTQA is centered around
Mg$^{2+}$, while the active site of 3CHY is built by residues only
(Asp12, Asp13, Met17, and Asp57).

It is possible that evolution follows several paths at the same time
and the question of whether evolution is divergent or convergent is
just ill-posed. To answer this, we still need more evidence to rule
out one scenario or another.

\section{Conclusion}

To conclude, we present a hierarchical model that attempts to explain
sequence conservation caused by the most basic and universal
evolutionary pressure in proteins to maintain stability. Using this
model, we show that separation of basic time scales (that constitute a
broad distribution with long tails) in evolution is a plausible
scenario for the sequence heterogeneity of structurally homologous
proteins. The two basic time scales are $\tau_o$ and $\tau\gg \tau_o$;
{\it (i)} at time scales, $\tau_o$, most mutations that occur in
protein sequences do not alter the protein's thermodynamically and/or
kinetically important sites and form families of homologous proteins;
{\it (ii)} at time scales $\tau\gg \tau_o$ mutations occur that would
alter several amino acids at the important sites of the proteins in
such a way that the properties of the proteins are not compromised. At
time scale $\tau$ the family of analogs is formed. Mutational rates,
directly computed during $Z$-score design, show agreement with the
conservatism profiles of the fold families.

The profile solution predicts sequence entropy reasonably well for the
majority, but not all, of amino acids. The amino acids that exhibit
considerably higher conservatism than predicted from stability
pressure alone are likely to be important for function and/or
folding. Comparison of the ``base-level'' stability conservatism
$S_{P}(k)$ with $S_{acr}(k)$ - actual conservatism profile of a
protein fold - allows one to identify functionally and kinetically
important amino acid residues and potentially gain specific insights
into folding and function of a protein.

Analysis of the correlation function confirms {\it (i)} the presence
of an intrinsic time scale, $\tau_o$, at which designed sequences are
similar and beyond which they differ strongly, {\it (ii)} the presence
of the conserved positions in the course of $Z$-score design. The
distributions of hamming distances between sequences reveal
``clustering'' of similar sequences (with low $Hd$) at short time
scales $\tau \sim \tau_o$ and disappearance of similarity at larger
time scales $\tau \gg \tau_o$. The above distributions are in accord
with the distribution of hamming distances in the families of
homologs, taken from the HSSP database, and with that of analogs,
taken from the FSSP database, correspondingly.

The proposed study offers a plausible explanation of the clustering of
structurally similar protein into families of homologs and
analogs. From the perspective of the proposed view of evolution, the
conservative amino acids appear as thermodynamically and kinetically
important centers, mutations of which result in other (possibly
strong) sequence modifications to preserve the physical properties of
the parental proteins. Such modifications result in a new family of
homologous proteins. In addition, the proposed model can be utilized
to search for the thermodynamically and kinetically important amino
acids in silica.

Evolution is an extremely complex phenomenon, driven by numerous
factors, such as history, preservation of function, folding kinetics
and stability of proteins in response to change in cell/body
environment. It is remarkable however, that our simple model was able
to qualitatively capture certain aspects of protein evolution without
any adjustable parameters (except for the contact definition threshold
and the empirical matrix of amino acid pairwise interactions).

In addition, our model offers an additional hint favoring a divergent
evolution. However, the question of whether evolution follows a
divergent or convergent path is yet to be resolved. Extensive
theoretical, phenomenological and experimental effort may bring
insight to this puzzle.

\section{Methods}

\subsection{Protein model}

We use the $C_{\beta}$ representation of proteins in which each pair
of amino acids is in contact if their $C_{\beta}$s ($C_{\alpha}$ in
the case of Gly) are within the distance 7.5\r A \cite{Jernigan96}.
We use Miyazawa-Jernigan (MJ) \cite{Miyazawa96} matrix of pair
potentials to represent the interaction between each pair of 20 amino
acids. The total potential energy of the protein can be written as
follows:
\be
E = \frac{1}{2} \sum_{i\neq j }^{N} U(\sigma_i, \sigma_j) \Delta_{ij} \, ,
\label{eq:en}
\ee
where N is the length of the protein, $\sigma_i$ is an amino acid at
the position $i = 1,\dots,N$. $U(\sigma_i, \sigma_j)$ is the
corresponding element of the MJ matrix of pairwise
interactions between amino acids $\sigma_i$ and $\sigma_j$.
$\Delta_{ij}$ is the element of the contact matrix, that is defined to
be 1 if contact between amino acids $i$ and $j$ exists (i.~e. the
distance between these amino acids in the native (ground) state is
smaller than 7.5 \r A), and 0, if the above contact does not exist:
\be
\Delta_{ij} \equiv \left\{ \begin{array}{ll}
                   1, &  |r_i^{NS}-r_j^{NS}|\le 7.5 \mbox{\r A}  \\
                   0, &  |r_i^{NS}-r_j^{NS}| >  7.5 \mbox{\r A}\, ,
                  \end{array} \label{eq:Delta}
\right.
\ee
where $r_i^{NS}$ is the position of the $i^{th}$ residue when the
protein is in the native conformation.

\subsection{The 6-letter potential}
\label{sseq:6l}

Due to the similarities in properties of the 20 types of amino acids,
one can classify these amino acids into 6 distinct groups: aliphatic
$\{AVLIMC\}$, aromatic $\{FWYH\}$, polar $\{STNQ\}$, positive
$\{KR\}$, negative $\{DE\}$, and special (reflecting their special
conformational properties) $\{GP\}$. We construct the potential of
interaction, $U_6(\hat{\sigma}_i, \hat{\sigma}_j)$, between the six
groups of amino acids, $\hat{\sigma}$, by computing the average
interaction between these groups, i.~e.
\be
U_6(\hat{\sigma}_i, \hat{\sigma}_j) = \frac{1}{N_{\hat{\sigma}_i} 
N_{\hat{\sigma}_j}} \sum_{\sigma_k \in \hat{\sigma}_i\, 
,\, \sigma_l \in \hat{\sigma}_j} U_{20}(\sigma_k, \sigma_l)\, ,
\label{eq:u6}
\ee
where $\sigma$ denotes amino acids in 20-letter representation and
$U_{20}(\sigma_k, \sigma_l)$ is the 20-letter matrix of interaction
MJ; $\hat{\sigma}$ denotes amino acids in 6-letter
representation. $N_{\hat{\sigma}}$ is the number of actual amino acids
of type $\hat{\sigma}$, e.~g. for the aliphatic group
$N_{\hat{\sigma}} = 6$. The 6-letter interaction potential for MJ
20-letter potential is given in Table~\ref{t:1}.

\subsection{The measure of the information context of the sequences}

In both the $Z$-score model and the profile solution, to study the
information context of the sequences, we compute the sequence entropy,
$S_X(k)$, at each position, $k$, of the sequence,
\be
S_X(k) = - \sum_{\sigma} P_X(\sigma_k)\ln P_X(\sigma_k)\, ,
\label{eq:s}
\ee
where $P_X(\sigma_k)$ is the probability that we observe an amino acid
$\sigma_k$ at the $k$th position. Subscript $X=Z$ or $P$ denotes the
$Z$-score model or profile solution correspondingly.  The summation is
taken over all possible values of $\sigma_k$.

The effect of switching to a 6-letter representation of amino acids
from the 20-letter representation on the sequence entropy, $S_6(k)$,
is that all values $S_6(k)$ are typically smaller than that of
$S_{20}(k)$. For a $M$-letter alphabet with all letters equally
represented, i.~e. $P_X(\sigma_k) = 1/M$, the entropy is equal to $\ln
M$. Thus, we expect that the difference between the typical values of
$S_{20}(k)$ and $S_6(k)$ is approximately $\ln(20/6)\approx 1.2$.  The
case when all letters of a $M$-letter alphabet are equally presented
corresponds to the maximal value of the entropy, i.~e.
\be
S_M(k)\leq \ln M\, .
\ee

\subsection{The entropy of the protein fold families}

Theoretical predictions from statistical-mechanical analysis can be
compared with data on real proteins. In order to determine
conservatism in real proteins we assume that the space of sequences that
fold into the same protein structure presents a two-tier system, where
homologous sequences are grouped into families and there is no
recognizable sequence homology between families despite the fact that
they fold into closely related structures
\cite{Rost97,Mirny99,Tiana00}.

Using the database of protein families with close sequence similarity
(HSSP database \cite{Dodge98}), we compute frequencies of amino acids
at each position, $k$, of aligned sequences, $P_m(\sigma_k)$, for a
given, $m$th, family of proteins. We average these frequencies {\it
across} all $N_s$ families sharing the same fold that are present in
FSSP database \cite{Holm93}:
\be
P_{acr}(\sigma_k) = \frac{1}{N_s}\sum_{m=1}^{N_s} P_m(\sigma_k)\, .
\ee
Next, we determine the sequence entropy, $S_{acr}(k)$, at each
position, $k$, of structurally aligned protein analogs:
\be
S_{acr}(k) = - \sum_{\sigma} P_{acr}(\sigma_k)\ln P_{acr}(\sigma_k) \, .
\label{eq:s1}
\ee

\section{Acknowledgments}
 
We thank R. S. Dokholyan for careful reading of the manuscript and
S. V. Buldyrev, A. V. Finkelstein, A. Yu. Grosberg, and L. A. Mirny
for helpful discussions. The profile solution was developed in
\cite{Dokholyan00xx} with L. A. Mirny.  NVD is supported by NIH
postdoctoral fellowship GM20251-01. EIS is supported by NIH grant
RO1-52126.

\appendix

\section{Determination of contact frequencies from homopolymer
conformations}
\label{ap:freq}

The estimation of frequencies is one of the key ingredients in protein
design. An alternative approach to that proposed above is to assume
that the set of conformational decoys is the set of all possible
random coil states of a homopolymer collapsed at the temperatures
below theta-point temperature, $T<T_{\theta}$ --- these are the states
that decoy random heteropolymers explore at the folding transition
temperature. Thus, we can determine the frequencies of contacts in an
ensemble of random heteropolymers by taking the time average of a
contact matrix element $\Delta_{ij}$ in the possible conformations of
a homopolymer at $T<T_{\theta}$.

To compute the frequencies of contacts for a homopolymer of length
$N$, we use discrete molecular dynamics simulations
\cite{Dokholyan00a,Zhou97a,Dokholyan98b}. We model a homopolymer by
$N$ beads on a string with the interaction distances scaled to 7.5\r
A. (see \cite{Dokholyan98b} for a detailed description of the model
and the algorithm). We run the simulation at the temperature
$T_{\theta}$ ($\epsilon$ parameter \cite{Dokholyan98b} is set to -1)
for $10^7$~time units\footnote{In the discrete molecular dynamics
algorithm, the time unit is the average time between subsequent
collisions}. After $10^7$ time units of simulations we compute the
frequency $f_{ij}$ of each of the $N(N-1)/2$ contacts in our
homopolymer.

\begin{figure}[hbt]
 \centerline{ \vbox{ \hbox{\epsfxsize=10.0cm
 \epsfbox{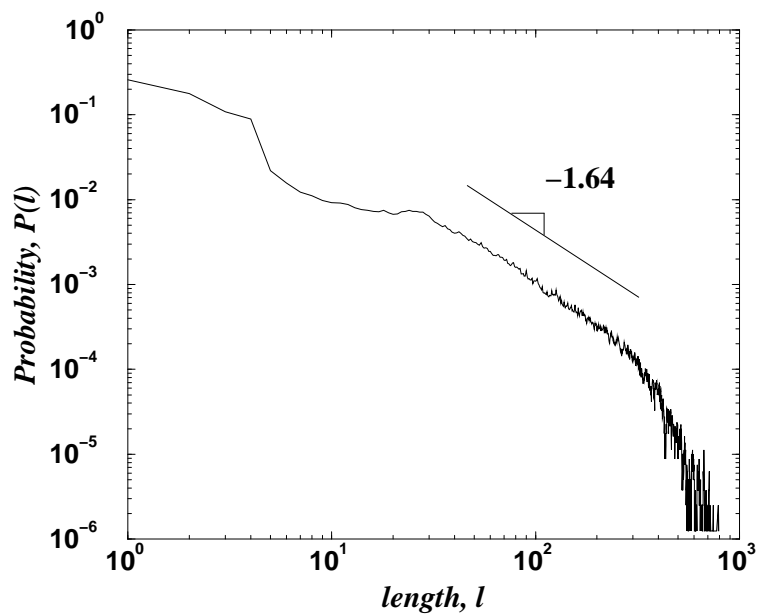}  }}}
 \vspace*{0.5cm}
\caption{Double-logarithmic plot of the distribution, $P(\ell)$, of
contacts of length $\ell=|i-j|$ in the ensemble of approximately
$10^3$ representative globular proteins in (PDB)
\protect\cite{Bernstein77,Abola87}. The contact between residues
positioned $i$th and $j$th along the protein chain is defined by the
Eq.~(\protect\ref{eq:Delta}) using $C_{\beta}$-representation of
proteins. The parallel line in the range of length $20 < \ell < 200$
indicates the power-law behavior of $P(\ell)$ in this region, $P(\ell)
\sim \ell^{-1.64}$. The region $5 < \ell < 20$ is specific
to proteins and has been discussed in detail in
\protect\cite{Berezovsky00}.}
\label{fig:freq}
\end{figure}

There are two principal drawbacks of the second method: {\it (1)} the
probability of occurrence of stable elements of the structure in
homopolymers resembling secondary structure in proteins is so low that
the distribution of contact lengths, $P(\ell)$, in homopolymer (not
shown) drastically differs from that shown for real proteins in
Fig.~\ref{fig:freq}. {\it (2)} The model of a homopolymer used in the
simulations strongly differs (e.g. in flexibility) from real
proteins. In fact, the problem of building an appropriate model for
chain flexibility is so important that small variations in it result
in drastically different kinetics from a realistic one
(e.g. appearance of the intermediate states)
\cite{Borreguero01,Ding01}. We find that both of these drawbacks make
the ``homopolymer'' approach of estimating the frequencies very
inefficient for existing protein models, so we omit it in our studies.

There are two strong advantages of this approach, which make it
worthwhile to explore it in the future, upon the availability of the
realistic protein models: {\it (i)} the possibility to generate a
large amount of decoy conformations and, thus, achieve statistically
highly significant contact frequency spectra; {\it (ii)} the
independence of the produced decoys from various database biases.

%\bibliographystyle{bibstyle}
%\bibliography{/home/dokh/biblio/ndbiblio}

\newpage

\include{tables}
\begin{table}[htb]
\begin{tabular}{l|c|cc|c|cc|c}
Fold&$N_s$&\multicolumn{2}{c}{Representative protein} &
    \multicolumn{4}{c}{Correlation coefficient, $r$}\\
    &    &PDB code \protect\cite{Bernstein77,Abola87}&$N$
    &$R(k)$ vs. $S_{acr}(k)$ 
    &$S_{P}(k)$ vs. $S_{acr}(k)$& $T_{sel}$  & $S_{P}(k)$ vs. $S_{acr}(k)$
($T_{sel}=0.25$)\\\hline
Ig  & 51 & $1TEN$ & 89   & 0.57 & 0.63 &  0.34  & 0.57\\
OB  & 18 & $1MJC$ & 69   & 0.67 & 0.69 &  0.19  & 0.69\\
R   & 166& $3CHY$ & 128  & 0.74 & 0.71 &  0.25  & 0.71\\
$\alpha/\beta$-P	
    &  29& $2ACY$ & 98   & 0.45 & 0.54 &  0.26  & 0.53\\
TIM &  49& $2EBN$ & 285  & 0.54 & 0.50 &  0.23  & 0.50\\
\end{tabular}
\vspace*{0.5cm}
\caption{The values of the correlation coefficient $r$ for the linear
regression of $S_{P}(k)$ and $R(k)$ versus $S_{acr}$ for Ig, OB, R,
$\alpha/\beta$-P, and TIM folds and the corresponding optimal values
of the temperature $T=T_{sel}$ for the $S_{P}(k)$ versus $S_{acr}$
linear regression. The last column corresponds to the correlation
coefficient for the studied folds at a fixed selective temperature
$T=0.25$. To obtain the rates of mutations, $R(k)$, we perform
$Z$-score design of the sequences for $t_d = 10^8$ Monte Carlo steps
at $T_{des}=0.25$.}
\label{t:2}
\end{table}
% for 1mjc at T=0.5, and td=10^7 r(R(k),S_acr(k))=0.77
%

\begin{table}[htb]
\begin{tabular}{l|cccccc}
%\multicolumn{7}{c}{6-letter MJ potential}\\\hline
  &    l  &    r  &    p  &    +  &    -  &    s  \\\hline
l & -0.31 & -0.39 & -0.22 &  0.01 & -0.41 & -0.12 \\
r & -0.39 & -0.27 & -0.32 & -0.02 & -0.28 & -0.12 \\
p & -0.22 & -0.32 & -0.41 & -0.25 &  0.07 & -0.29 \\
+ &  0.01 & -0.02 & -0.25 & -0.10 & -0.18 & -0.18 \\
- & -0.41 & -0.28 &  0.07 & -0.18 &  0.01 & -0.05 \\
s & -0.12 & -0.12 & -0.29 & -0.18 & -0.05 &  0.04 \\
\end{tabular}
\vspace*{0.5cm}
\caption{A 6-letter potential derived (see Section~\ref{sseq:6l}) for
MJ 20-letter potential. The symbols ``l'', ``r'', ``p'', ``+'', ``-''
and ``s'' denote 6 distinct corresponding groups of amino acids:
aliphatic $\{AVLIMC\}$, aromatic $\{FWYH\}$, polar $\{STNQ\}$,
positively charged $\{KR\}$, negatively charged $\{DE\}$, and special
(reflecting their special conformational properties) $\{GP\}$.}
\label{t:1}
\end{table}
\clearpage

\end{document}